\documentclass[%
preprint,
 amsmath,amssymb,
 aps,
]{revtex4-2}

\usepackage{graphicx}
\usepackage{dcolumn}
\usepackage{bm}
\usepackage{hyperref}

\usepackage{siunitx}
\usepackage[version=4]{mhchem}
\usepackage{xcolor}

\let \Gamma \varGamma
\let \Psi \varPsi
\let \Phi \varPhi

\newcommand{\pderd}[2]{\frac{\mathrm{d} #1}{\mathrm{d} #2}}
\newcommand{\cH}{c_{\ce{H+}}}

\newcommand{\cOH}{c_{\ce{OH-}}}
\renewcommand{\DH}{D^\prime_{\ce{H+}}}
\newcommand{\DOH}{D^\prime_{\ce{OH-}}}
\newcommand{\jac}{j_\mathrm{ac}}
\newcommand{\Kw}{K_w}
\newcommand{\TKw}{\widetilde{K}_w}
\newcommand{\Keq}{K}
\newcommand{\TK}{\widetilde{K}}
\newcommand{\Cchr}{C_0}

\newcommand{\sech}{\mathrm{sech}}
\newcommand{\Cs}{C_S}
\newcommand{\cs}{c_S}
\newcommand{\Ndl}{\mathcal{N}}

\begin{document}


\title{Theory of ion and water transport in electron-conducting membrane pores with pH-dependent chemical charge}
\underline{}
\author{L. Zhang}
\email{li.zhang@wetsus.nl}
\author{P. M. Biesheuvel}%
\affiliation{%
 Wetsus, European Centre of Excellence for Sustainable Water Technology, Oostergoweg 9, 8911 MA Leeuwarden, The Netherlands}%


\author{I. I. Ryzhkov}
\email{rii@icm.krasn.ru}
\affiliation{
 Institute of Computational Modelling SB RAS, Federal Research Center KSC SB RAS, Akademgorodok 50-44, 660036 Krasnoyarsk, Russia}%
 \affiliation{Siberian Federal University, Svobodny 79, 660041 Krasnoyarsk, Russia}



\begin{abstract}
In this work, we develop an extended uniform potential (UP) model for a membrane nanopore by including two different charging mechanisms of the pore walls, namely by electronic charge and by chemical charge. These two charging mechanisms will generally occur in polymeric membranes with conducting agents, or membranes made of conducting materials like carbon nanotubes with surface ionizable groups. The electronic charge redistributes along the pore in response to the gradient of electric potential in the pore, while the chemical charge depends on the local pH via a Langmuir-type isotherm. The extended UP model shows good agreement with experimental data for membrane potential measured at zero current condition. When both types of charge are present, the ratio of the electronic charge to the chemical charge can be characterized by the dimensionless number of surface groups and the dimensionless capacitance of the dielectric Stern layer. The performance of the membrane pore in converting osmotic energy from a salt concentration difference into electrical power can be improved by tuning the electronic charge.
\end{abstract}

\maketitle

\section{Introduction}
Many porous membranes bear charge on the surface of their pore walls. Often this charge is due to the polymeric or inorganic membrane material, which can (de)protonate to leave a charged surface group~\cite{Behrens2001,Coronell2008}. The chemical nature of the charge is determined by the equilibrium between the surface groups and solution, which typically has a strong dependence on pH.
In other materials where the pore walls are conductive, the membrane can be charged electronically~\cite{Nishizawa1995,Lebedev2018}.
Sometimes, both chemical charge and electronic charge can exist at the same time: either by introducing ionizable charged groups on conducting materials, or adding conducting agents, such as carbon nanotubes, in polymeric membranes~\cite{Biesheuvel2016,Fornasiero2008,Jiang2010,Zhang2019}. The electrostatic effect of these surface charge plays a significant role in modulating transport of ionic species through the membrane, and has been engineered to provide new approaches for energy conversion~\cite{Guo2010,Siria2017}, desalination~\cite{Kim2010a}, separation~\cite{Zhang2019a}, fabrication of ion field-effect transistors~\cite{Daiguji2004,Jiang2010} and mimicking biological cell membranes~\cite{Guan2012}.

The nature of the electronic and the chemical charge is very different. The total electronic charge in the pore walls will be a conserved quantity when there is no electron supply or leakage via the external circuit or due to a Faradaic reaction.
Still, the electronic charge will redistribute over the pore length to ensure that the electronic potential is the same everywhere, i.e., to achieve an equipotential pore wall surface. 
This redistribution can lead to regions of negative and positive charge, i.e., the membrane pore is polarized, similar to the polarization of conducting particles~\cite{Roche2014} and porous carbons~\cite{Rubin2016}.
The chemical charge, however, has a very different origin and  only depends on the composition of solution and the pore wall chemistry. In a typical scenario, the hydronium ion (proton) is the most common charge-determining ion and thus the chemical charge usually has a strong dependence on the local pH in the pore. Dynamics of surface (de)protonation is usually much faster compared with dynamics of ion transport, and thus instantaneous chemical equilibrium is commonly assumed for the generation of chemical charge (charging dynamics may be important in certain cases, for example, \cite{Werkhoven2018}). For a steady-state transport problem, even if the chemical equilibrium on the surface is only slowly established, we can still describe the chemical charge by an equilibrium adsorption model, such as the Langmuir isotherm.

In general, both types of surface charge are nonuniformly distributed along the pore. To ensure equipotential on the pore wall, the electronic charge is redistributed to compensate for the non-uniform potential distribution in the solution of the pore interior. Meanwhile, the chemical charge varies with the pH in the solution. The distribution of the surface charge will strongly influence the transport of ions and the performance of the membrane. Some simple scenarios of varying surface charge have been studied, for example, a step-wise distribution for asymmetric membranes \cite{Takagi1986} and a linear distribution \cite{Manzanares1991}. However, this effect is still far from being well-understood, manifested by the prevailing assumption of constant surface charge in modeling ion transport in the membrane, both within the two-dimensional space charge (SC) model~\cite{Gross1968, Fair1971, Sasidhar1981}, and the one-dimensional uniform potential (UP) model~\cite{Wang1997,Bowen2002,Kim2010}. In this work, we extend the UP model by incorporating both the \emph{electronic charge} (induced by the electric field in the pore solution) and the \emph{chemical charge} (determined by local pH) to investigate the effect of surface charge on transport of ions through the membrane.

\begin{figure}
    \centering
    \includegraphics[scale = 1]{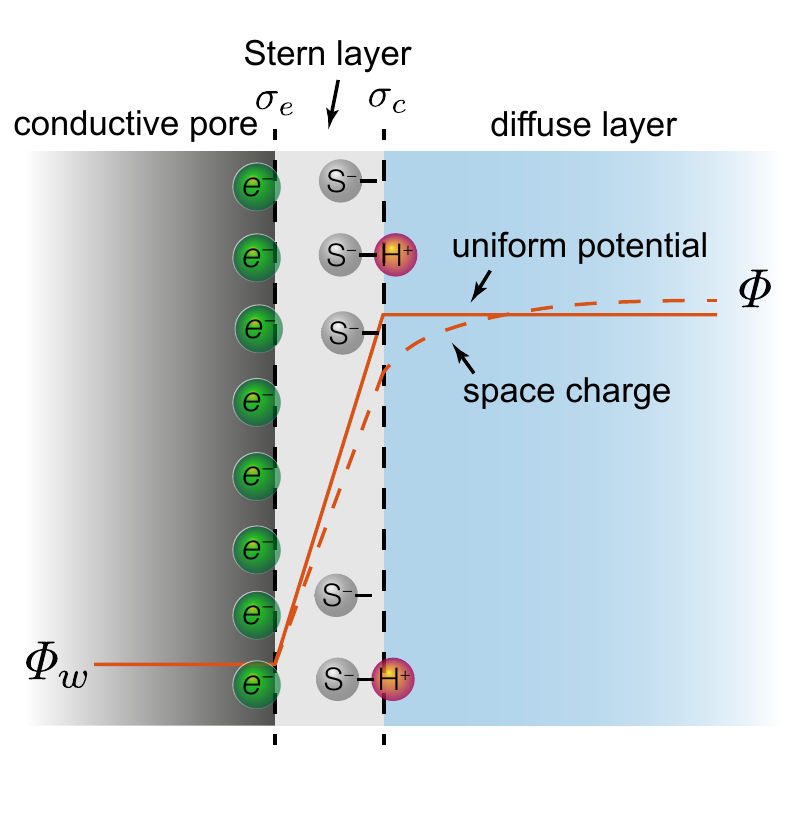}
    \caption{Structure of the electrical double layer with electronic and chemical charge on the pore wall and ionic charge in the solution.}
    \label{fig:charging}
\end{figure}

The distribution of surface charge along the membrane pore strongly affects membrane transport properties, such as the membrane potential, which is the potential difference between two electrolyte solutions with different salt concentrations separated by the membrane. In the present work, we consider the membrane potential at a condition of zero ionic current. 
The measurement of membrane potential is useful in characterizing the ionic permselectivity of ion exchange membranes \cite{Yaroslavtsev2003}, interpreting the measurement of potentiometric ion sensing \cite{Bobacka2008}, evaluating the maximum power that can be generated in reverse electrodialysis (RED) \cite{Post2007,Tedesco2016}, and modulating cellular activities as a key biophysical signal in biological cell membranes \cite{Yang2013}. 
Conventionally, the membrane potential is ascribed to the two Donnan potentials due to the electrical double layers at the membrane-solution interfaces and the diffusion potential within the membrane due to different mobilities of ions \cite{Lakshminarayanaiah1969,Strathmann2011}. Recently, \textcite{Ryzhkov2017} reported a new mechanism for the generation of membrane potential in polarizable conductive membranes via the induced electronic charge. The re-distribution of electronic charge enhances the membrane potential when there is a difference in mobilities between the cation and anion. In the present work, we demonstrate that our extended UP model can quantitatively capture this effect for small pores by comparing with the experimental data and the results of the full two-dimensional space charge model. Moreover, we show that the variation of chemical charge due to gradients in the proton concentration also contributes to a potential difference within the membrane, and gives rise to a decrease of membrane potential at large salt concentration ratios, which has been observed in previous work \cite{Kim2010, Galama2016}, but has not yet been explained.
In addition, for cases with both electronic and chemical charge, we propose a dimensionless parameter to quantify the ratio of the electronic charge to the chemical charge and study the performance of the membrane in generating osmotic power from a salt concentration difference by a reverse electrodialysis process.

Our paper is organized as follows: the framework of our model is introduced in Section 2, the main results and analysis are presented in Section 3, where we discuss first the case of only electronic charge, then the case of only chemical charge, and finally the general case where both types of charge play a role. Section 4 concludes the work.

\section{Theoretical Model}
Let us consider a membrane separating two reservoirs with aqueous solutions of the same monovalent and symmetric (1:1) electrolyte of concentrations $C_h$ and $C_l$, respectively ($C_h \geqslant C_l$). The reservoirs are maintained at equal hydrostatic pressures. The membrane is modelled as an array of pore channels of length $L_p$ and characteristic pore size $H_p$. Depending on the cross-sectional geometry, $H_p$ corresponds to the radius for a cylindrical pore, or the width of a planar channel for a slit pore. We assume that at each position along the pore, the Debye length $\lambda_D$ is of the same order of, or larger than, the characteristic size $H_p$.
The Debye length follows from $ \lambda_D = \sqrt{\varepsilon\varepsilon_0 R_g T/2F^2 C_0}$, where $\varepsilon\varepsilon_0$ is the dielectric constant of the solution, $R_g$ is the ideal gas constant, $T$ is temperature, $F$ is the Faraday constant and $C_0$ is the characteristic concentration in the problem. 
In this case, electric potential $\Phi$, cation concentration $C_+$ and anion concentration $C_-$ as well as hydrostatic pressure $P$ can be assumed uniform in any cross--section of the pore, so they are functions of axial coordinate only. This approach is known as the uniform potential (UP) model or the ''fine capillary pore model'' and is sometimes called the Teorell-Meyer-Sievers (TMS) model~\cite{Teorell1953}, though the TMS model does not include fluid flow. The UP model is a simplification of the space charge model, which solves the 2D Navier--Stokes (NS) and Poisson--Nernst--Planck (PNP) equations using the approaches of virtual variables, i.e., hypothetical variables of a solution equilibrated with adjacent differential elements of the membrane ~\cite{Spiegler1966,Gross1968,Fair1971,Sasidhar1981,Peters2016}.

As shown in Fig.~\ref{fig:charging}, we consider two charging mechanisms at the surface of the membrane pores: the electronic charge $\sigma_e$ [\SI{}{C/m^2}] in the electron-conducting pore wall, and the chemical charge $\sigma_c$ [\SI{}{C/m^2}] originating from the deprotonation of surface groups $\ce{S}$ (e.g., hydroxylic or carboxylic groups) according to a reversible reaction
\begin{equation}
\ce{SH} \rightleftarrows \ce{S^-} + \ce{H^+} \,. \nonumber
\end{equation}
The equilibrium is characterized by the dissociation constant $\Keq$ and the maximum number of ionizable sites $N$ per m$^2$,
\begin{subequations}\label{eq:Langmuir_KN}
\begin{align}
\Keq & = {[\ce{S^{-}]} \{\ce{H^+}\}}\big/{[{\ce{SH}}]}\,, \label{eq:Langmuir_K}\\
N & = [{\ce{SH}}] + [{\ce{S^{-}}}]\,, \label{eq:Langmuir_N}
\end{align}
\end{subequations}
where $[{\ce{SH}}]$ and $[{\ce{S^{-}}}]$ are the surface concentrations of non--ionized and ionized surface groups, while $\{\ce{H^+}\}$ is the proton concentration near the pore surface, which is also denoted as $C_{\ce{H^+}}$ below.
Combining Eqs.~\eqref{eq:Langmuir_KN} leads to the density of chemical surface charge described by the well-known Langmuir 1--pK adsorption isotherm
\begin{gather}
\sigma_c = -e \, [{\ce{S^{-}}}] = -e N \frac{1}{1 + 10^{\mathrm{pK - pH}}}\,, \label{Eq1}
\end{gather}
where $e$ is the elementary charge, $\mathrm{pK} = -\log_{10} \Keq$ and $\mathrm{pH} = -\log_{10}\,{C_{\ce{H^+}}}$ with $\Keq$ and $C_{\ce{H^+}}$ in the unit of $\SI{}{M}$, i.e., $\SI{}{mol/L}$.

It is further assumed that the chemical charge is located at the interface between a dielectric layer, which is referred to as the Stern layer, and  the diffuse layer, and thus separated from the electronic charge in the conductive pore wall.
Similar models including both ionic and electronic charging processes have been developed for capacitive deionization in~\cite{Dykstra2017}, for electrofluidic gating of chemically reactive surface in~\cite{Jiang2010,Jiang2011,Mei2016} and  for oxidized metal or semiconductive oxides in \cite{Duval2001, Duval2002}.

The electronic charge induced at the pore wall is given by a linear relation considering that no charge exists inside the dielectric Stern layer,
\begin{equation}
    \sigma_e= \Cs \left(\Phi_w - \Phi_s \right)\,,  
\end{equation}
where $\Phi_w$ is the potential in the electronic conducting pore wall, and $\Phi_s$ is the potential at the Stern plane, which is the intersection of the Stern layer and the diffuse layer. In the UP model, $\Phi_s$ coincides with the potential throughout the aqueous phase in the pore ($\Phi_s = \Phi$), see Fig.~\ref{fig:charging},
while $\Cs$ is the capacitance of the Stern layer, which is expected to depend on the permittivity $\varepsilon_s$, thickness $\delta$, and geometry of the Stern layer. 
The total surface charge entering the electroneutrality condition is $\sigma = \sigma_c + \sigma_e$, and the corresponding volumetric density of the wall charge is given by $2\sigma / F H_p$. If there are no chemical surface groups, i.e., $N=0$ and thus $\sigma_c=0$, the surface can only be charged electronically so that $\sigma = \sigma_e$, while in the other limit, the electronic charge is set to zero along the pore and $\sigma = \sigma_c$. In addition, the present model treats the chemical charge as a smeared-out surface charge, and we do not consider its point-like character. This is a reasonable approximation as long as the surface charge density is relatively large, but may fail for low charge density such as biological lipid membranes, where a point-charge model for the chemical charge has to be employed \cite{Nelson1975}.


To proceed further, let us introduce dimensionless variables by choosing the characteristic scales for length $L_p$, for potential $\Phi_T = R_g T/F$, for concentration and volumetric charge density $\Cchr$, for pressure $\Cchr R_g T$, for fluid velocity $D/L_p$ and for ion fluxes $D \Cchr/L_p$.
Here $D = \sqrt{D_+ D_-}$ is the average diffusion coefficient, where $D_+$ and $D_-$ are the cation and anion diffusion coefficients. The chemical and electronic charge densities in dimensionless form are written as
\begin{subequations}
\label{eq:dml_X}
\begin{align}
X_c & = \frac{2\sigma_c}{\Cchr F H_p} = -\frac{\Ndl}{ 1 + 10^{\mathrm{pK-pH}}}\,, \label{eq:dml_Xc} \\
X_e & = \frac{2\sigma_e}{\Cchr F H_p} = 
\cs \left(\phi_w - \phi\right), \label{eq:dml_Xe}
\end{align}
\end{subequations}
where $\Ndl = 2 N (\Cchr N_A H_p)^{-1}$ is dimensionless density of surface group sites, $\cs = 2 \Cs R_g T (\Cchr F^2 H_p)^{-1}$ is dimensionless Stern layer capacitance and $\phi_w$ and $\phi$ are dimensionless potentials corresponding to $\Phi_w$ and $\Phi$. The averaged electronic charge along the pore length is defined as $\overline{X}_e=\int_0^1 X_e\, \mathrm{d}z$. When the membrane is not charged externally by injecting or withdrawing electrons to or from the conductive membrane pore walls, one has $\overline{X}_e=0$.
However, even then the local value $X_e(z)$ can be non--zero since electrons are redistributed along the surface in order to ensure equipotential in the pore wall~\cite{Ryzhkov2017,Ryzhkov2018}. 
The volumetric density of wall charge is $X=X_c + X_e$, which is opposite in sign to the density of ionic charges $c_+ - c_-$ in the pore because of total charge neutrality
\begin{gather}
    c_+ - c_- + X = 0, \label{eq:neutral}
\end{gather}
where we have assumed that the concentrations of \ce{H^+} and \ce{OH^-} are much lower than the concentrations of cations and anions arising from salt dissociation, so that their presence is not taken into account in the electroneutrality condition \eqref{eq:neutral} and neither in the total ionic flux and ionic current.

The equations of the UP model for transport of water and ions at steady-state with unequal diffusion coefficients and variable surface charge are given by \cite{Catalano2018}
\begin{subequations}
\label{eq:up_uneqD}
\begin{align}
& u = - \frac1{\Theta \alpha} \pderd{p}{z} + \frac{X}{\Theta \alpha} \pderd{\phi}{z}\,, \label{eq:up_u} \\
& j_\mathrm{s} = c_T \, u - \cosh(\xi) \,\pderd{c_T}{z} - \sinh(\xi)\, \pderd{X}{z} + \big(\cosh(\xi) \, X + \sinh(\xi) \,c_T \big) \pderd{\phi}{z}\,, \label{eq:c_flux} \\
& j_\mathrm{ch} = - X \, u + \sinh(\xi) \, \pderd{c_T}{z} + \cosh(\xi)\, \pderd{X}{z}  - \big( \cosh(\xi)\, c_T + \sinh(\xi) \, X \big)\pderd{\phi}{z}\,. \label{eq:charge_flux}
\end{align}
\end{subequations}
where $u$ is the fluid velocity along the pore direction $z$, $j_s =j_+ + j_-$ is the total solute flux of cations and anions, $j_\mathrm{ch} = j_+ - j_-$ is the flux of the ionic charge, $c_T = c_+ + c_-$ is the total concentration of cations and anions, $p$ is the dimensionless hydrostatic pressure, $\alpha = \mu D \, (\Cchr R_g T H_p^2)^{-1}$ is the dimensionless viscosity parameter with $\mu$ the fluid viscosity, and $\xi = \ln(\sqrt{D_-/D_+})$ is a factor accounting for the effect of unequal diffusion coefficients. A shape factor $\Theta$ is introduced to account for different cross-sectional geometries: $\Theta = 8$ for cylindrical pore and $\Theta = 12$ for slit-shaped pore. Note that $u$, $j_\mathrm{s}$ and $j_\mathrm{ch}$ are constants along the pore in the current model, while for a dynamic problem, $u$ and $j_\mathrm{ch}$ are still invariants due to the continuity of fluid flow and electric current, but $j_\mathrm{s}$ will vary along the pore. Furthermore, we neglect the possible relevance of upstream and downstream diffusion boundary layers~\cite{Catalano2018}.

To account for the variation of the chemical charge determined by the local pH in the solution, we need to supplement equations \eqref{eq:up_uneqD} by the transport equations of \ce{H+} and \ce{OH-} ions including advection, diffusion, and electromigration~\cite{Zhang2018},
\begin{subequations}
\begin{align}
j_{\ce{H+}} & = \cH \, u - \DH \left( \pderd{\cH}{z} + \cH \pderd{\phi}{z} \right), \label{eq:H_flux} \\
j_{\ce{OH-}} & = \cOH \, u - \DOH \left( \pderd{\cOH}{z} - \cOH \pderd{\phi}{z} \right),
\end{align}
\end{subequations}
where $j_{\ce{H+}}$ and $j_{\ce{OH-}}$ are the dimensionless fluxes of \ce{H+} and \ce{OH-} scaled by $D \Cchr/H_p $, 
$\DH = D_{\ce{H+}}/D$ and $\DOH = D_{\ce{OH-}}/D$ are the dimensionless diffusion coefficients of \ce{H+} and \ce{OH-}, respectively ($D_{\ce{H+}} = \SI{9.32e-9}{m^2/s}$, $D_{\ce{OH^-}} = \SI{5.26e-9}{m^2/s}$~\cite{Newman2004}). At steady state, the flux of \ce{H^+} and \ce{OH^-} into the surface due to ionization is zero. Mass conservation of \ce{H+} and \ce{OH-} requires that the difference between the fluxes, i.e., the acidity flux, $\jac = j_{\ce{H^+}} - j_{\ce{OH^-}}$, is constant. We replace the concentration of \ce{OH-} according to the water dissociation equilibrium $\cH\,\cOH = \TKw$ where $\TKw = \Kw/\Cchr^2$ is the dimensionless equilibrium constant. It leads to 
\begin{equation}
    \jac = \Big( \cH - \frac{\TKw}{\cH} \Big) \, u - \Big( \DH + \frac{\DOH \TKw }{\cH^2} \Big) \Big( \pderd{\cH}{z} + \cH \pderd{\phi}{z}\Big)\,. \label{eq:ac_flux}
\end{equation}
In the current model, the resulting variation of pH along the pore described by Eq.~\eqref{eq:ac_flux} determines the chemical surface charge via the Langmuir isotherm \eqref{Eq1}, and this chemical charge in turn couples back into various transport properties. However, except for this back-coupling, there is no other direct effect of the transport of \ce{H^+} and \ce{OH^-}. For typical conditions with moderate pH values ($\sim 5 - 9 $), this is a valid approximation as the salt concentration is much higher than that of the proton and hydroxide ions. Besides, for acidic conditions when the proton is dominant ($\cH \gg \sqrt{\TKw}$), equation \eqref{eq:ac_flux} reduces to \eqref{eq:H_flux}.

Now we specify the boundary conditions of the problem. The membrane separates two reservoirs with dimensionless salt concentrations $c_h$ and $c_l$, as well as hydrostatic pressures $p_h$ and $p_l$, which are assumed to be the same. The concentrations of \ce{H^+} ions in the reservoirs are set by specifying the pH values. The membrane potential $\Delta \phi$ is defined as the potential on the low concentration side minus that on the high concentration side. Due to the large aspect ratio of the pore geometry ($L_p \gg H_p$), the membrane-solution interfaces are treated using the classical Donnan model~\cite{Tian2015}, in which the following boundary conditions are set at the two pore ends:
\begin{subequations}
\label{eq:bc1}
\begin{align}
p{(z)} & = p_\mathrm{res} - 2c_\mathrm{res} + c_{T}{(z)}, \label{eq:bc_p} \\
c_{T}{(z)} & = 2 c_\mathrm{res} \cosh\left(\Delta\phi_\mathrm{Donnan}{(z)}\right), \label{eq:bc_c} \\
\cH{(z)} & = c_{\ce{H+},\mathrm{res}} \exp\left(-\Delta\phi_\mathrm{Donnan}{(z)}\right) \quad \text{or} \quad \mathrm{pH}{(z)} = \mathrm{pH}_\mathrm{res} + \Delta\phi_\mathrm{Donnan}{(z)}/\ln{10}, \label{eq:bc_cH}
\end{align}
\end{subequations}
where the pore entrance ($z=0$) is connected to the high-concentration reservoir ($\mathrm{res} = h$) and the pore exit ($z=1$) is connected to the low-concentration reservoir ($\mathrm{res} = l$). The Donnan concentration jumps are described by conditions \eqref{eq:bc_c} and \eqref{eq:bc_cH}, while condition \eqref{eq:bc_p} corresponds to the osmotic pressure jump. The Donnan potential $\Delta\phi_\mathrm{Donnan}$ is defined as the potential within the pore minus the potential outside at the membrane-reservoir interface. By inserting $c_\pm{(z)} = c_\mathrm{res} \exp\left(\mp\Delta\phi_\mathrm{Donnan}{(z)}\right)$ into the charge neutrality condition \eqref{eq:neutral}, we have
\begin{equation}
    X{(z)} = 2 c_\mathrm{res} \, \sinh\left(\Delta\phi_\mathrm{Donnan}{(z)}\right). \label{eq:bc_X1}
\end{equation}
Further inserting the expressions for the chemical and electronic charge \eqref{eq:dml_X} and boundary conditions \eqref{eq:bc_cH} leads to
\begin{equation}
-\frac{\Ndl}{1 + 10^{\mathrm{pK-pH}_\mathrm{res}}\exp\left(-\Delta\phi_\mathrm{Donnan}{(z)}\right)}  + \cs \left(\phi_w - \phi{(z)}\right) = 2c_\mathrm{res}\, \sinh\left(\Delta\phi_\mathrm{Donnan}{(z)}\right). \label{eq:phiD}
\end{equation}
After setting the reference potential at high concentration reservoir to zero for simplicity, Eq.~\eqref{eq:phiD} can be solved with respect to $\Delta\phi_\mathrm{Donnan}(z=0) = \phi(0)$. The potential $\phi(1) $ at the other end is found by solving the transport equations in the pore, while $\Delta \phi_\mathrm{Donnan}(z=1)$ is again found from the solution of \eqref{eq:phiD}. The membrane potential $\Delta \phi$ is calculated as the potential variation across the inner coordinates of the membrane $\phi{(z=1)}-\phi{(z=0)}$, plus the Donnan potential jumps at each end, i.e., adding $\Delta\phi_\mathrm{Donnan}(z=0)$ and subtracting $\Delta\phi_\mathrm{Donnan}(z=1)$.

\textcolor{black}{To solve the system, we substitute $\mathrm{d}c_T/\mathrm{d}z$ in \eqref{eq:charge_flux} using \eqref{eq:c_flux}, express $\mathrm{d}X/\mathrm{d}z$ as the sum of the electronic and chemical parts by \eqref{eq:dml_X} and further replace $\mathrm{d} \cH /\mathrm{d} z$ using \eqref{eq:ac_flux}. Then we obtain an ordinary differential equation for $\mathrm{d}\phi/\mathrm{d}z$,}
\begin{equation}
    f_3 \pderd{\phi}{z} = \sinh(\xi)\, (c_T \,u - j_s) - \cosh(\xi) \, X u - \cosh(\xi) \, j_\mathrm{ch} + f_1 \,f_2, \label{eq:dphidz}
\end{equation}
where
\begin{equation}
     f_1(\cH) = \pderd{X_c}{\cH} = \frac{\Ndl \TK}{\left(\cH + \TK \right)^2}, \quad f_2(\cH)  = \frac{\left(\cH - \TKw/\cH \right) u - \jac}{\DH +\DOH \TKw/\cH^2}, \quad f_3(c_T, \cH)  = c_T + \cs + f_1 \cH,
\end{equation}
and $\TK = \Keq/\Cchr$ is the dimensionless equilibrium constant for surface deprotonation. The terms on the right hand side of \eqref{eq:dphidz} represent the potential difference in the pore caused by diffusion of salt ions with unequal mobilities, solvent flow (streaming potential), Ohm's resistance and variation of surface charge, respectively. If the diffusion coefficients of ions are the same, i.e., $\xi =0$, the potential for unequal diffusion will vanish, while if the fluid velocity $u$ goes to zero, the streaming potential caused by the flow of solute in the diffuse layer, which is not electroneutral, will disappear. 
Formally, the scale factor $f_3$ can be seen as an effective conductivity consisting of three parts representing contributions from solute concentration, electronic charging and chemical charging. The electronic charging process increases the total conductivity by redistributing the electronic charge in the conducting wall to generate a reverse electrical field and reduce the potential drop in the pore, while the chemical charging process modifies the potential drop through transport of proton, which modulates the chemical surface charge. The role of chemical charging is more complicated as it not only adds to the conductivity, but also affects the potential difference by the term $f_1 f_2$.

Then, we rearrange the equations of the UP model \eqref{eq:up_uneqD} and the flux equation \eqref{eq:ac_flux} for \ce{H^+} to a set of ordinary differential equations
\begin{subequations}
\label{eq:sol_sys}
\begin{align}
\pderd{p}{z} & = - 8\alpha\, u + X \pderd{\phi}{z}, \label{eq:dpdz} \\
\pderd{c_T}{z} & = \sech(\xi)\, (c_T u - j_s) - \tanh(\xi) \, f_1\,f_2 + \Big( X + \tanh(\xi)\, f_3 \Big) \pderd{\phi}{z}, \label{eq:dcdz} \\
\pderd{\cH}{z} & = f_2 - \cH  \pderd{\phi}{z}. \label{eq:dcHdz}
\end{align}
\end{subequations}
Note that $\mathrm{d}X/\mathrm{d}z$ has been expressed through $\mathrm{d}\phi/\mathrm{d}z$ and $\mathrm{d}{\cH}/\mathrm{d}{z}$ using $X=X_c + X_e$ and Eq.~\eqref{eq:dml_X}, and $\mathrm{d}{\cH}/\mathrm{d}{z}$ is further replaced using Eq.~\eqref{eq:dcHdz}. The shooting method is used to solve the problem by integrating \eqref{eq:sol_sys} from $z=0$ to $z=1$ and matching the hydrostatic pressure, salt concentration and pH in the reservoir connected to the pore exit. 
\begin{figure}[t]
\begin{minipage}[m]{0.5\textwidth} 
    \centering
    \includegraphics[scale = 0.7]{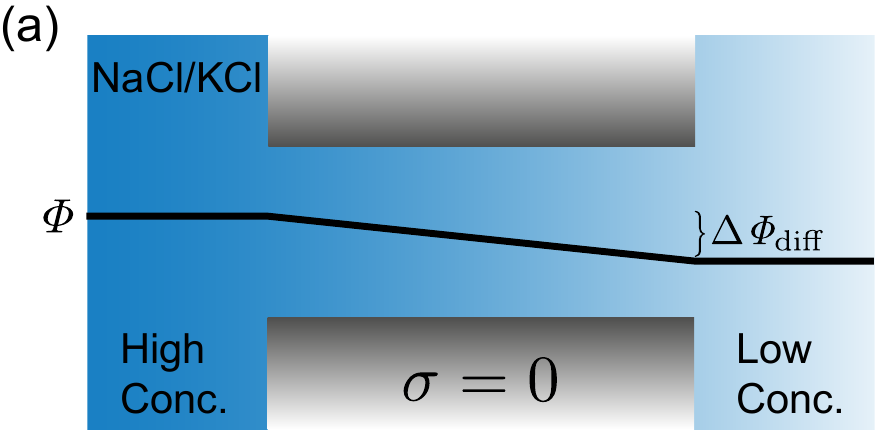}
\end{minipage}
\begin{minipage}[m]{0.5\textwidth} 
    \centering
    \includegraphics[scale = 0.7]{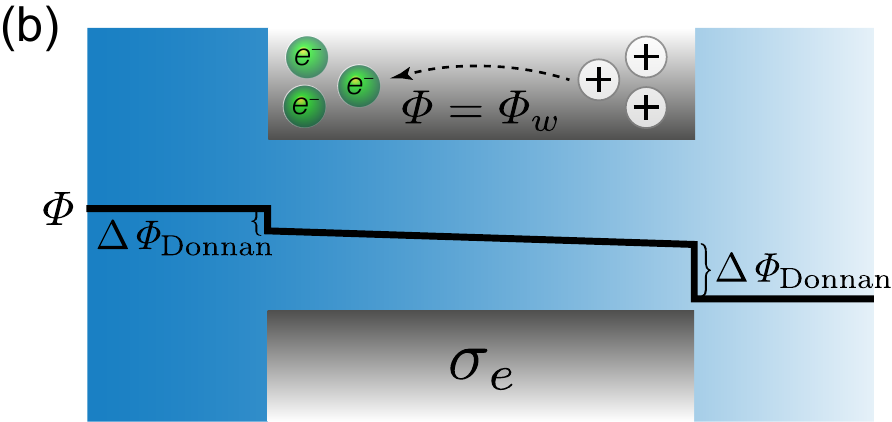}
\end{minipage}%
\begin{minipage}[m]{0.5\textwidth} 
    \centering
    \includegraphics[scale = 0.7]{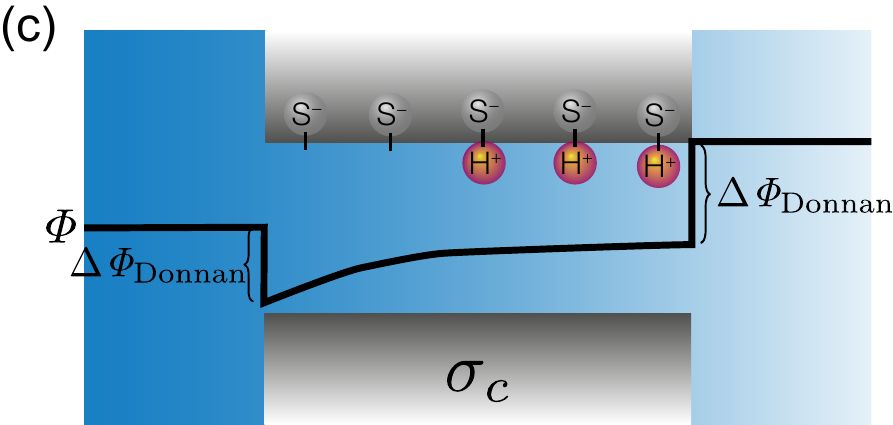}
\end{minipage}
    \caption{Typical profiles of electrical potential for zero current condition in a membrane pore (a) without surface charge, (b) with only electronic charge, and (c) with only chemical charge. The ions diffuse from the left high-concentration reservoir to the right low-concentration one. $\Delta\Phi_\mathrm{diff}$ is the diffusion potential of an uncharged pore with non-conductive surface, and $\Delta\Phi_\mathrm{Donnan}$ is the Donnan potential at the membrane-reservoir interface.}
    \label{fig:sketch}
\end{figure}

\section{Results and Discussion}

In this section, we analyze the predictions of the steady-state uniform potential model for different charging conditions on the pore walls. The aqueous \ce{NaCl} and \ce{KCl} solutions are considered with diffusion coefficients $D_{\ce{Na+}} = \SI{1.33e-9}{m^2/s}$, $D_{\ce{K+}} = \SI{1.96e-9}{m^2/s}$ and $D_{\ce{Cl-}} = \SI{2.03e-9}{m^2/s}$~\cite{Newman2004}. First in section~\ref{sec:electro}, we show the results for the case of only electronic charge demonstrated in Fig.~\ref{fig:sketch}b and compare them with the 2D space charge model as well as the experimental data. Next in section~\ref{sec:chemical}, we study the case of only chemical charge determined by local pH (Fig.~\ref{fig:sketch}c). After that, in section~\ref{sec:combine}, we discuss more general scenarios when both electronic and chemical charge play a role. Most results are for a condition of zero applied electric current through the membrane $j_\mathrm{ch} = 0$, and the resulting membrane potential is presented as the main outcome. In section \ref{sec:energy_generation}, we also present results for a non-zero electric current to analyze the electric power production from a salt concentration difference, i.e., "osmotic power" or "blue energy"~\cite{Siria2017}.

\subsection{Case of only electronic charge} \label{sec:electro}
Let us start with the case where only electronic charge is on the pore walls. Throughout this section, we assume that the total electronic charge $\overline{X_e}$ is zero, i.e., no extra electrons are injected or withdrawn from the membrane.

If the mobilities or diffusivities of the cation and anion are different, there is a spontaneous electrical field generated to ensure local electroneutrality when the ions diffuse from the high-concentration side to the low-concentration side. For \ce{NaCl}/\ce{KCl}, anions move faster than cations, so that the electrical potential drops along the membrane to reduce the speed of the anions and raise the speed of the cations (Fig.~\ref{fig:sketch}a). For an uncharged, non-conductive membrane, this diffusion potential across the membrane can be calculated as \cite{Ryzhkov2017}
\begin{equation}
    \Delta \Phi_\mathrm{diff} = \phi_T \frac{D_+ - D_-}{D_+ + D_-} \ln\frac{C_h}{C_l}  = \phi_T \frac{\exp(-2\xi)-1}{\exp(-2\xi)+1} \ln\frac{C_h}{C_l}.
    \label{eq:diff_phi}
\end{equation}

If the membrane pore wall is electron conducting, the spontaneous electrical field that develops in the pore interior will exert an electrical force on the electrons in the pore, which will re-distribute to guarantee equipotential in the conducting pore wall. For the case illustrated in Fig.~\ref{fig:sketch}b, the diffusion potential generates an electrical force along the membrane and pushes the electrons in the membrane from the low-concentration side to the high-concentration side. This leads to a negative(positive) surface charge near the high-concentration(low-concentration) side. Consequently, the Donnan potentials at both ends, acting in the same direction, enlarge the total potential drop across the membrane. This enhancement effect has been reported for C-Nafen membrane \cite{Ryzhkov2017}, which was prepared from alumina nanofibers covered by a conductive carbon layer.

\begin{figure*}[!htb]
    \centering
    \begin{minipage}[t]{0.5\textwidth}
    \includegraphics[scale=0.7]{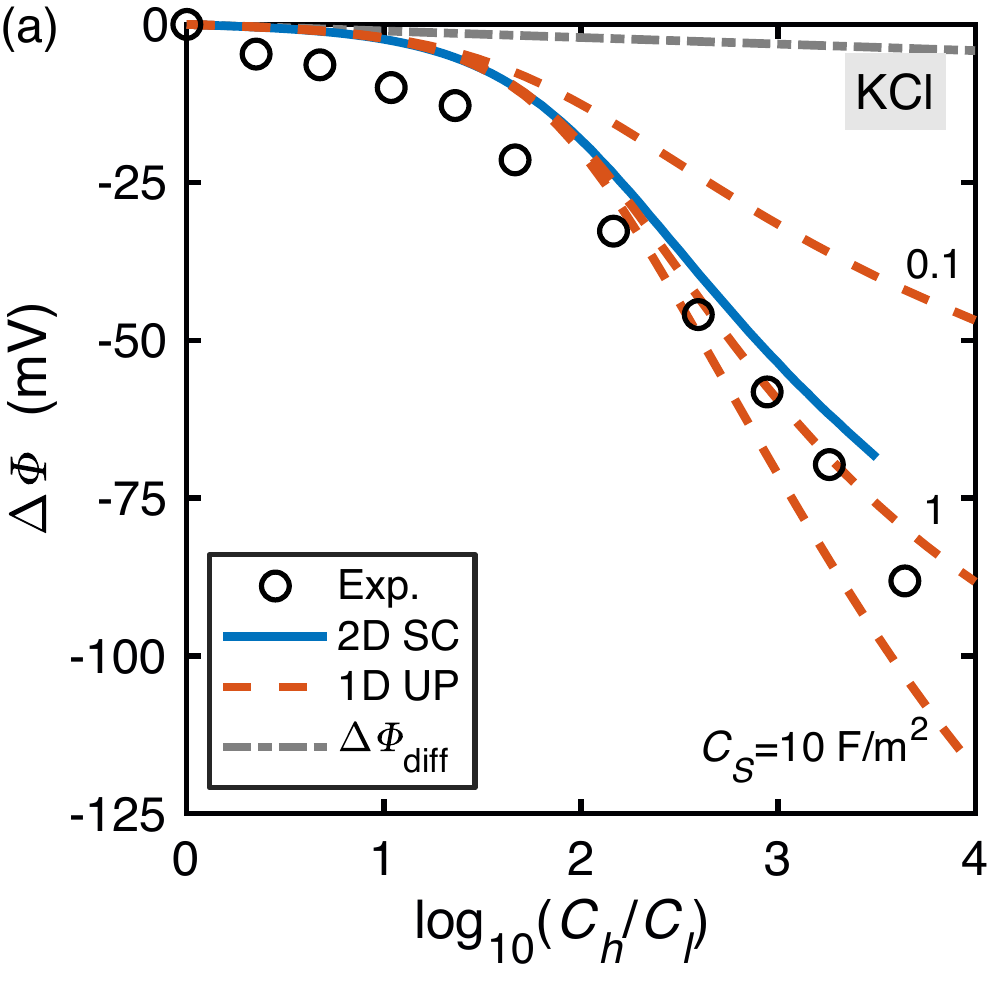}
    \end{minipage}%
    \begin{minipage}[t]{0.5\textwidth}
    \includegraphics[scale=0.7]{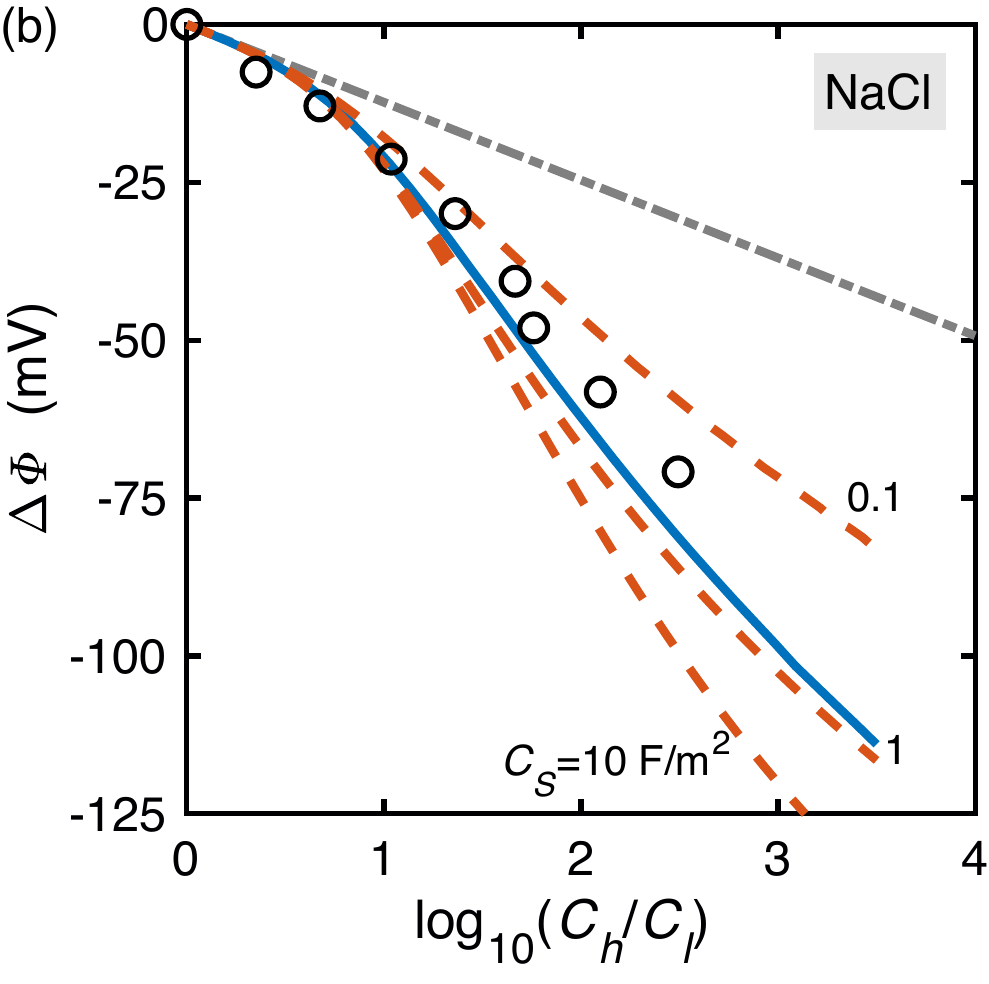}
    \end{minipage}    
    \caption{Membrane potential of conductive C-Nafen membrane for different concentration ratios in (a) KCl and (b) NaCl aqueous solution. Data points and results of 2D space charge model without Stern layer shown in solid blue lines are from \textcite{Ryzhkov2017}. Dashed red lines are results of 1D uniform potential model with $\Cs = 0.1, 1, \SI{10}{F/m^2}$, and dash-dotted lines are the diffusion potentials in uncharged pores with non-conductive surface. $H_p = \SI{8}{nm}$, $C_l = \SI{0.1}{mM}$ for \ce{KCl} and $\SI{1}{mM}$ for \ce{NaCl}.}
    \label{fig:Dphi_Xe}
\end{figure*}

\begin{figure}[!htb]
    \begin{minipage}[t]{0.5\textwidth}
    \includegraphics[scale=0.7]{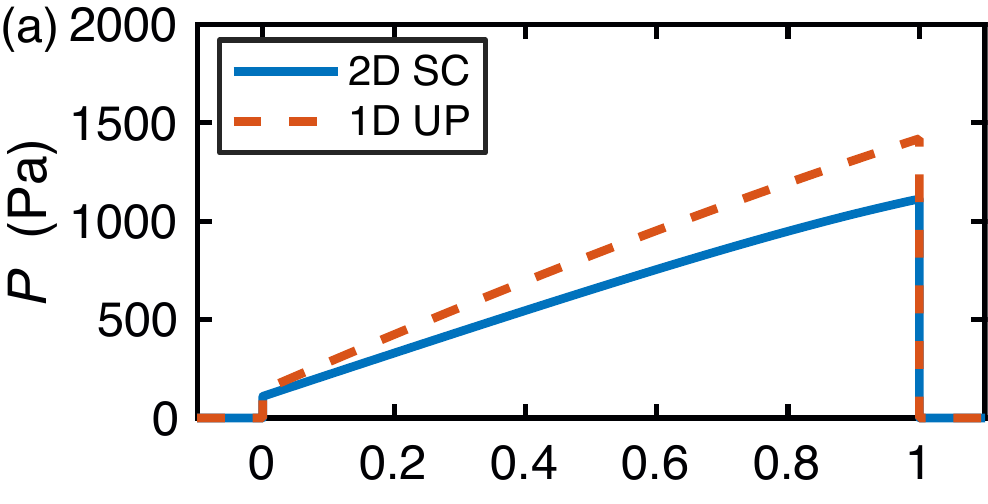}
    \end{minipage}%
    \begin{minipage}[t]{0.5\textwidth}
    \includegraphics[scale=0.7]{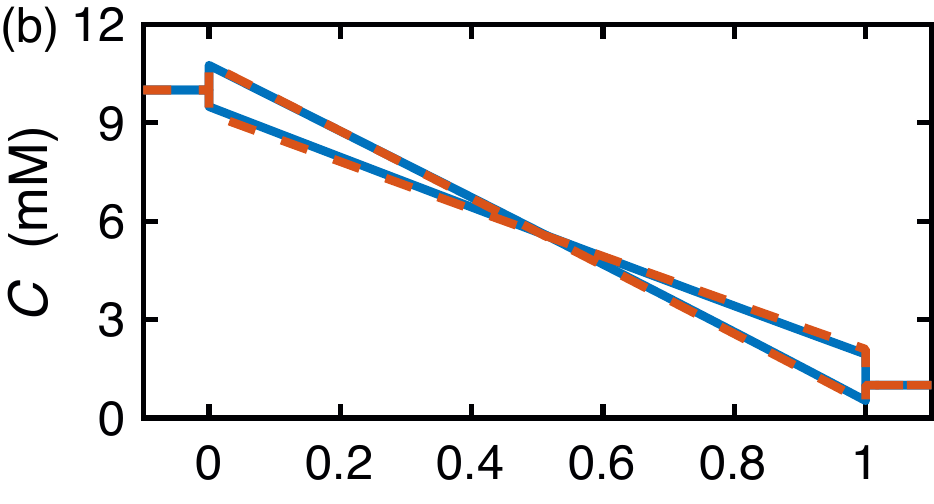}
    \end{minipage}
    \begin{minipage}[t]{0.5\textwidth}
    \includegraphics[scale=0.7]{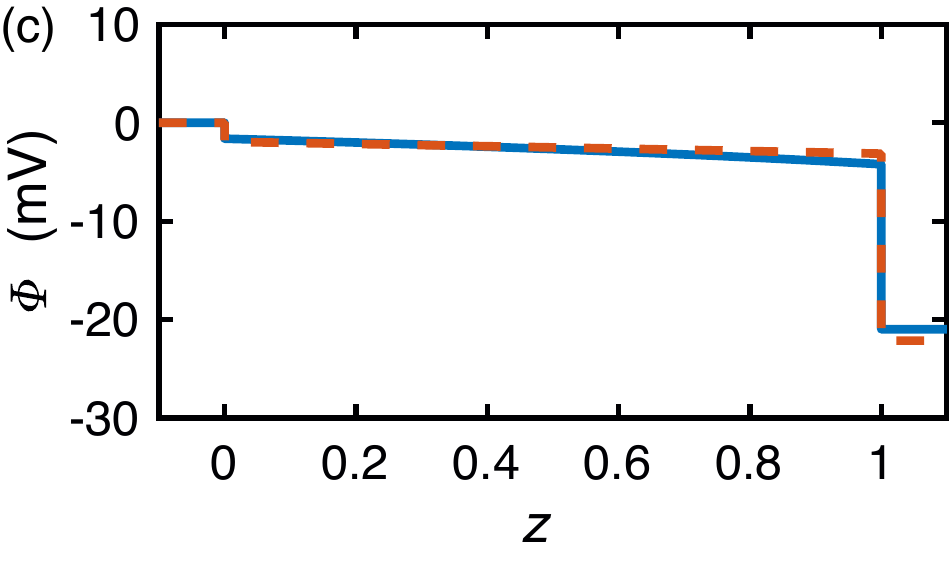}
    \end{minipage}%
    \begin{minipage}[t]{0.5\textwidth}
    \includegraphics[scale=0.7]{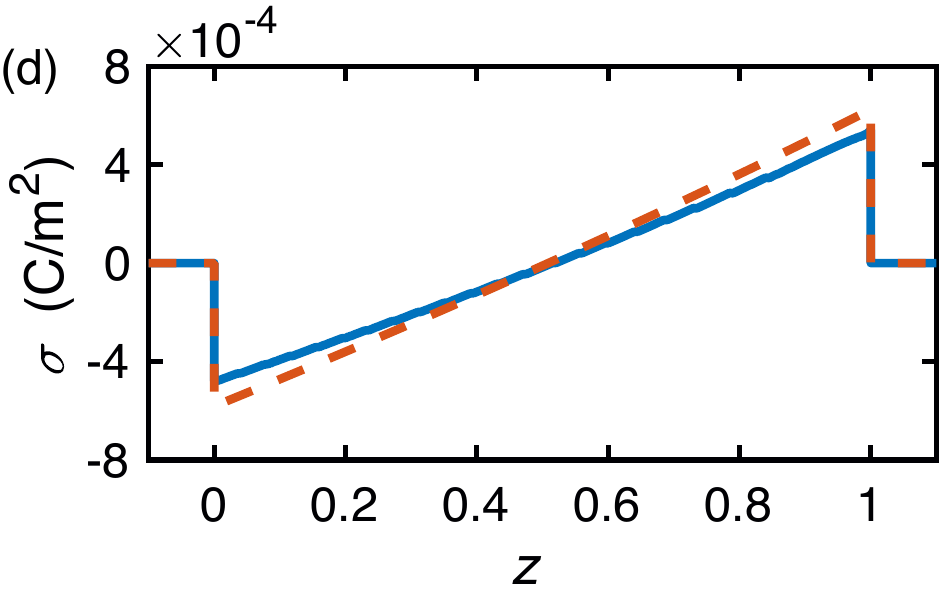}
    \end{minipage}     
    \caption{Profiles of (a) pressure, (b) cation and anion concentration, (c) electrical potential and (d) surface charge density in a $\SI{8}{nm}$-radius pore filled with \ce{NaCl} at zero current. $C_h = \SI{10}{mM}$ and $C_l = \SI{1}{mM}$. Solid lines are results of the 2D space charge model without Stern layer, and dashed lines are results of the 1D uniform potential model with $\Cs = \SI{1}{F/m^2}$.}
    \label{fig:profile_Xe}
\end{figure}

Figure~\ref{fig:Dphi_Xe} shows the comparison of measured membrane potential with model predictions based on the 1D UP model and 2D SC model for a $\SI{8}{nm}$-radius pore. Note that even for \ce{KCl} with a minor difference in diffusion coefficients, which are usually ignored in some theoretical studies, the membrane potential can be enhanced to a few times or even dozens of times with the increase of the concentration ratio. The 2D SC model shown as blue lines has no Stern layer~\cite{Ryzhkov2017}, or equivalently, assuming $\Cs \rightarrow \infty$. Nevertheless, the Stern layer is necessary in the 1D UP model to relate the charge density with the potential difference $\Phi_w - \Phi$ across the interface. Three different values of the Stern layer capacitance $\Cs$ are used in the UP model shown as dashed lines in Fig.~\ref{fig:Dphi_Xe}. A good agreement is obtained for $\Cs = \SI{1}{F/m^2}$, and this value will be used throughout the paper. Theoretically, the capacitance $\Cs$ in the UP model can be pushed to a large value to make a direct comparison with the SC model, and the difference between them is ascribed to the one-dimensional assumption for a pore with finite radius. Figure~\ref{fig:profile_Xe} further shows a comparison between the UP model and the SC model of the profiles of pressure, concentration, electrical potential and surface charge density in a pore with \ce{NaCl} and radius $H_p = \SI{8}{nm}$, $C_h = \SI{10}{mM}$, $C_l = \SI{1}{mM}$. The two have quantitative agreement and it is found that this agreement is reasonably well for pore sizes smaller than $\SI{10}{nm}$ with the concentration up to $\SI{1}{M}$. In this sense, the capacitance in the UP model can be seen as a good fitting parameter, which mitigates some error from the one-dimensional assumption.

\begin{figure}[!htb]
    \begin{minipage}[t]{0.3\textwidth}
    \includegraphics[scale=0.6]{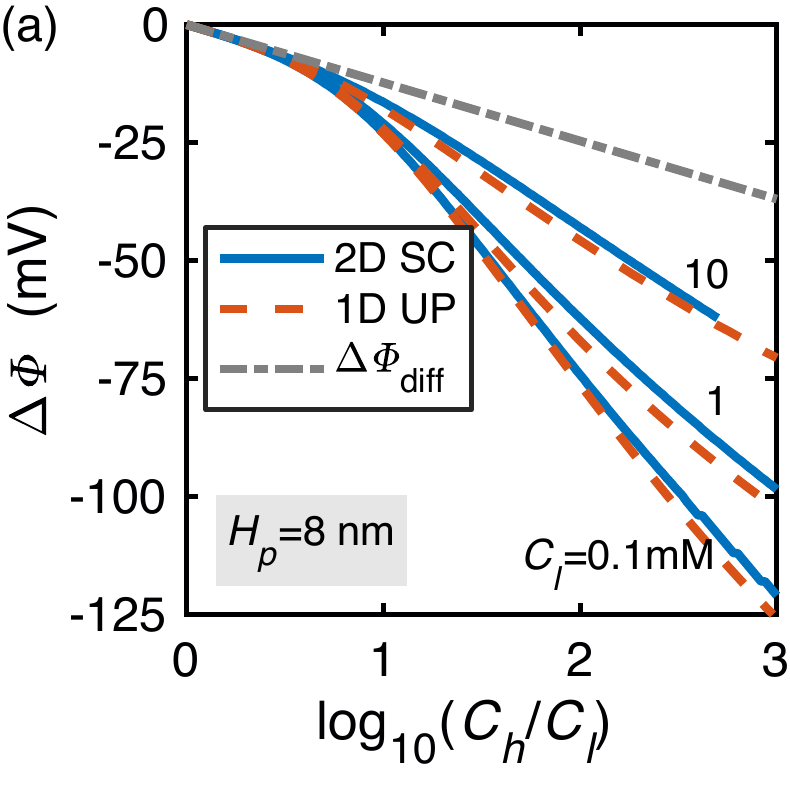}
    \end{minipage}
    \begin{minipage}[t]{0.33\textwidth}
    \includegraphics[scale=0.6]{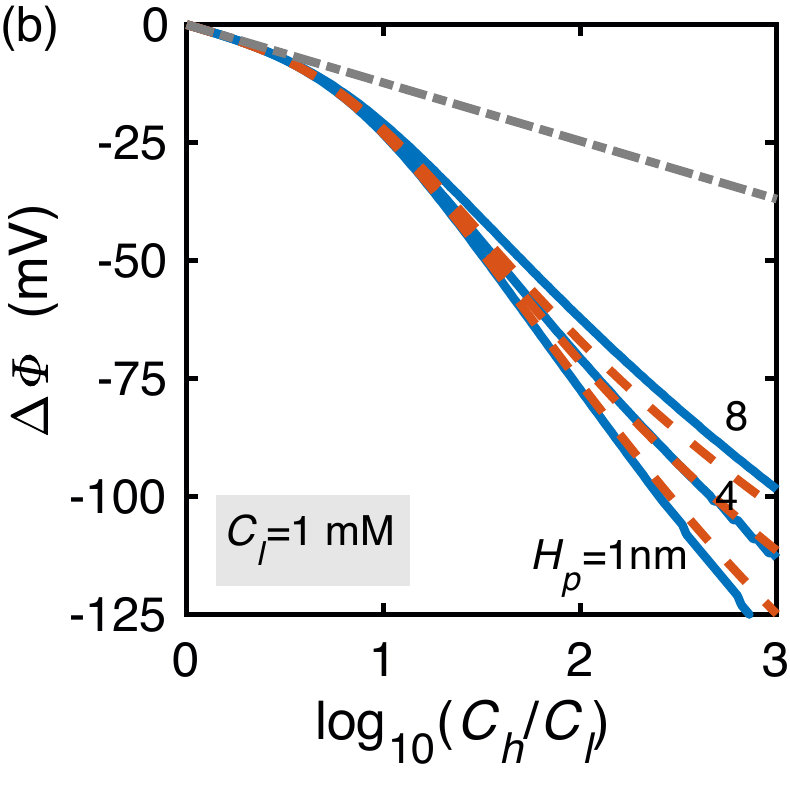}
    \end{minipage}
    \begin{minipage}[t]{0.33\textwidth}
    \includegraphics[scale=0.6]{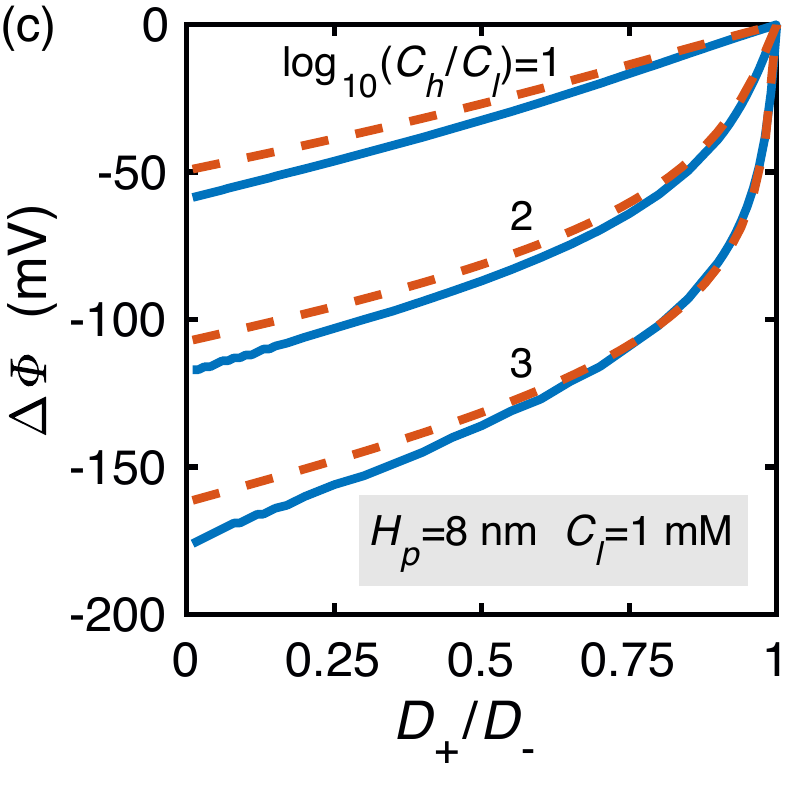}
    \end{minipage}
    \caption{Parametric study of (a) concentration ($H_p = \SI{8}{nm}$), (b) pore radius ($C_l = \SI{1}{mM}$),  and (c) diffusion coefficients ($H_p = \SI{8}{nm}, C_l = \SI{0.1}{mM}$) on membrane potential in cylindrical nanopore filled with \ce{NaCl}. Dashed lines are results of uniform potential model with $\Cs = \SI{1}{F/m^2}$, solid lines are that of the 2D space charge model without Stern layer, and dash-dotted lines are the diffusion potentials in uncharged pores with non-conductive surface.}
    \label{fig:Dphi_Rp_Cl_D}
\end{figure}

Figure~\ref{fig:Dphi_Rp_Cl_D} shows the effect of concentration, pore radius, and diffusion coefficients on the membrane potential. For all the conditions, the UP model agrees well with the SC model. When the reservoir concentration or the pore radius increases, the Donnan potential at the membrane-reservoir interface decreases according to
\begin{equation}\label{eq:Donnan_for_sigmae}
    \Delta \Phi_\mathrm{Donnan} = \sinh^{-1} \Big( \frac{\sigma}{F H_p C_\mathrm{res}}\Big),
\end{equation}
where $\sigma$ is the surface charge density at the pore end and $C_\mathrm{res}$ is the corresponding reservoir concentration. Therefore, the magnitude of the membrane potential declines as the enhancement effect via the Donnan potential drops. When the ratio between the diffusivities $D_+/D_-$ reduces, the diffusion potential becomes stronger and it leads to an increase of the electronic surface charge density $\sigma_e$. From equation \eqref{eq:Donnan_for_sigmae}, it is clear that the Donnan potential, and thus the membrane potential will be enlarged.

\subsection{Case of only chemical charge} \label{sec:chemical}
In this section, we consider the case where the membrane is charged only by chemical groups. The charge regulation by pH is considered by incorporating the transport equation of proton and hydroxide ions described by Eq. \eqref{eq:ac_flux}. It has been shown in recent studies that this regulation mechanism has significant effect on ionic conductance~\cite{Ma2015} and in electro-osmotic hysteresis~\cite{Zhang2018}. 

\begin{figure}[htb]
    \centering
    \begin{minipage}[t]{0.5\textwidth}
    \includegraphics[scale=0.7]{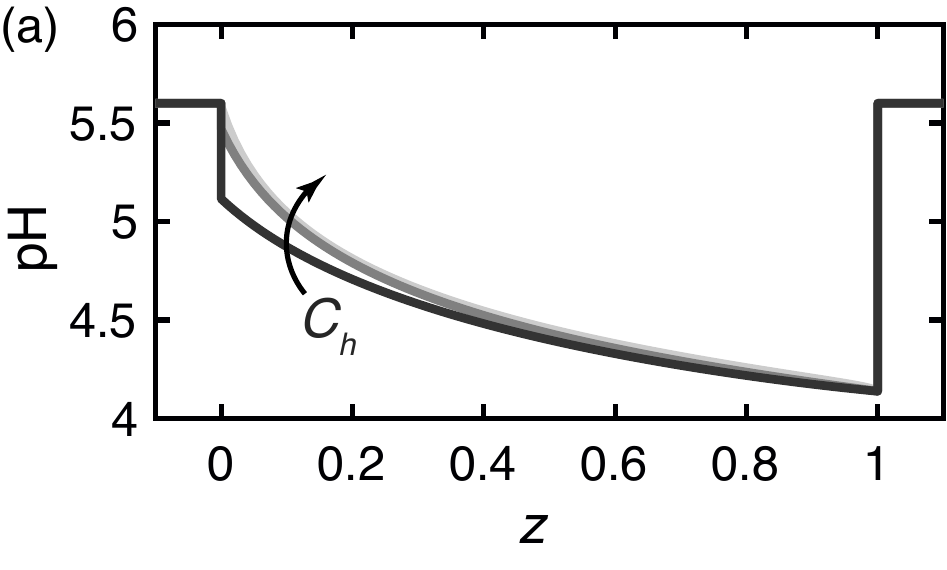}
    \end{minipage}%
    \begin{minipage}[t]{0.5\textwidth}
    \includegraphics[scale=0.7]{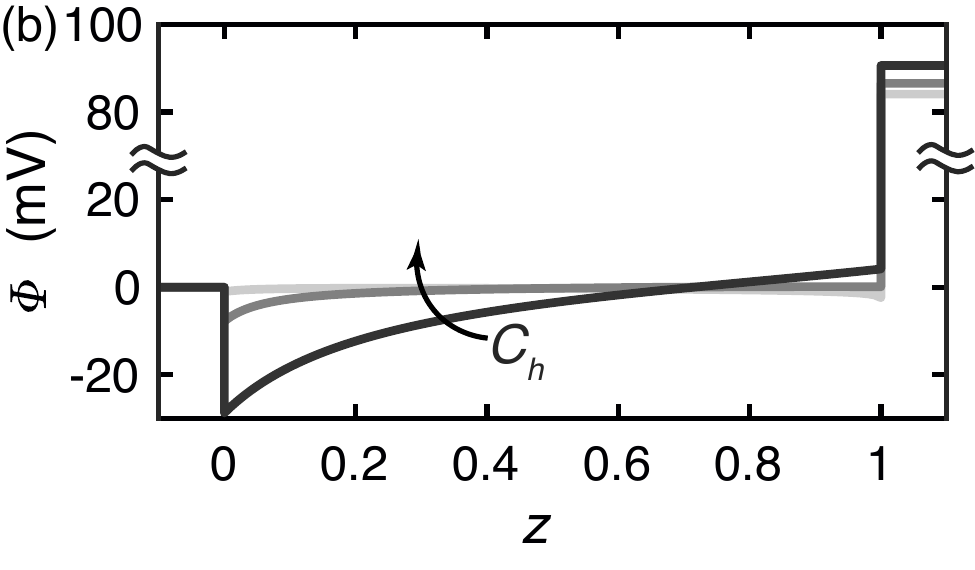}
    \end{minipage}      
    \caption{Profile of (a) pH and (b) electrical potential of a $\SI{4}{nm}$-width slit channel with chemical surface charge. $C_l=\SI{0.1}{mM}$ and $C_h$ increases from $10, 100$ to $\SI{1000}{mM}$.}
    \label{fig:profile_Xc}
\end{figure}

The pH in both reservoirs is kept the same in all cases. However, due to the Donnan potentials at each end, the proton concentration(pH) increases(decreases) at the low-concentration end (see Fig.~\ref{fig:profile_Xc}a), lowering the surface charge density via combining with the surface groups. The variation of surface charge results in an electrical field acting opposite to the concentration gradient, so the potential increases along the pore, as depicted in Fig.~\ref{fig:sketch}c. This corresponds to the term $f_1 f_2/f_3$ in equation \eqref{eq:dphidz}. If the surface charge is not dependent on pH, this term vanishes and it reduces to the constant surface charge model.

\begin{figure*}[htb]
    \centering
    \begin{minipage}[t]{0.5\textwidth}
    \includegraphics[scale=0.7]{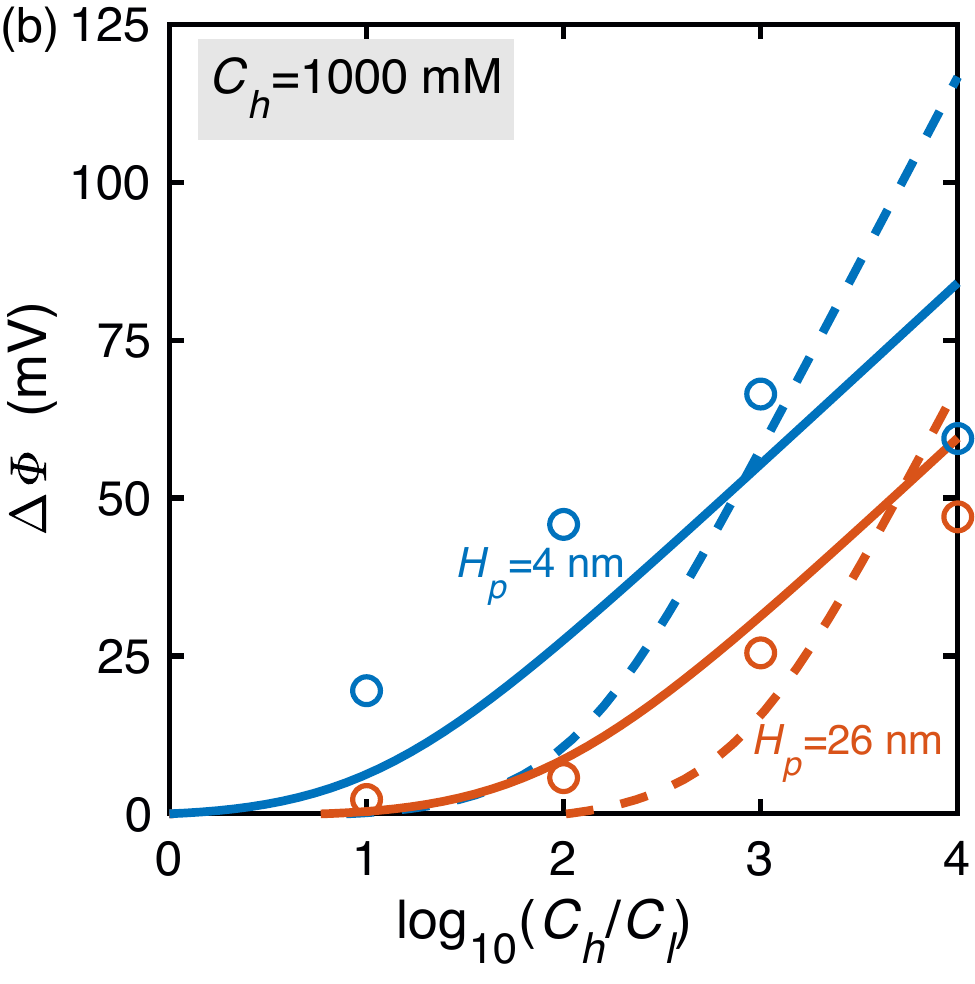}
    \end{minipage}%
    \begin{minipage}[t]{0.5\textwidth}
    \includegraphics[scale=0.7]{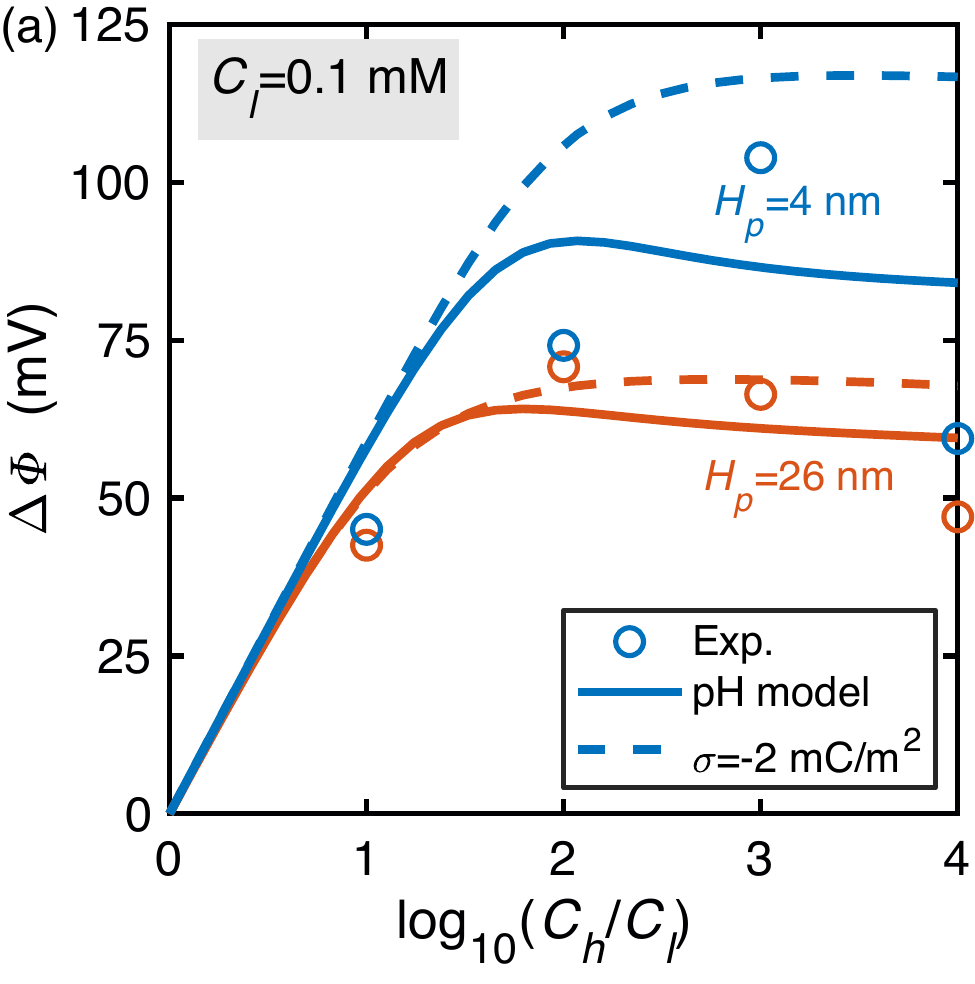}
    \end{minipage}   
    \caption{Membrane potential across $\SI{4}{nm}$- and $\SI{26}{nm}$-width slit channels filled with \ce{KCl}. Experimental data points are from \textcite{Kim2010} at zero current. Solid lines are results of the uniform potential model with pH-dependent chemical charge (silica surface with $\mathrm{pK} = 7.5$ and $N = \SI{8}{nm^{-2}}$, $\mathrm{pH}=5.6$ for both reservoirs), while dashed lines are results of the constant surface charge model ($\sigma = \SI{-2}{mC/m^2}$).}
    \label{fig:Dphi_Xc}
\end{figure*}

Figure~\ref{fig:Dphi_Xc} shows the membrane potential across $\SI{4}{nm}$ and $\SI{26}{nm}$ nanoslit channels for different concentration ratios, with the lower concentration fixed in panel (a) and the higher concentration fixed in panel (b). In general, as the concentration ratio increases, the membrane potential increases mainly due to the contribution of the Donnan potential. However, as the concentration ratio exceeds around 100 with a fixed lower concentration (Fig.~\ref{fig:Dphi_Xc}a), the Donnan potential at the high-concentration end drops to around zero because the surface charge is much lower than the ionic charge carried by cation and anions. Therefore, the membrane potential reaches a plateau with further increase of the concentration ratio. This scenario is predicted by the constant surface charge model with $\sigma = \SI{-2}{mC/m^2}$.

However, the experimental data in Fig.~\ref{fig:Dphi_Xc}a shows a systematic decrease of the membrane potential when the concentration ratio becomes relatively large for both $\SI{4}{nm}$ and $\SI{26}{nm}$ channels. This decrease of membrane potential can be partly captured by the current model, in which the proton transport and the variation of chemical charge are considered. The reason behind this is that the contribution of potential increase owing to the variation of chemical charge diminishes as the conductance of the pore increases with the reservoir concentration. Specifically, when the concentration $C_h$ is relatively low, the co-ions are mostly repelled from the pore and the conductance is controlled by the surface charge density. Nevertheless, when the concentration $C_h$ becomes higher, both the counter-ions and co-ions can go into the pore and the conductivity in this case is controlled by the ionic charge, thus leading to a dramatic increase of the total conductance. This weakens the potential increase owing to the variation of the chemical charge (Fig.~\ref{fig:profile_Xc}).

The current UP model achieves a better agreement with the experimental data compared with the constant surface charge model without any fitting procedure. Parameters used in the 1-pK Langmuir isotherm, namely, the equilibrium constant of surface deprotonation and the number of surface sites, are well-constrained values reported in the literature \cite{Behrens2001}. One may wonder if the assumption that protons and hydroxide ions do not contribute to the charge density and flux leads to the unexpected decrease of membrane potential in the current model. Considering the relatively high salt concentration and medium pH, this assumption should be reasonable. In fact, we performed further simulations using a full multi-component model including protons and hydroxide ions in charge density and flux. This extended model gives quantitatively consistent results with the current model and predicts a decreasing membrane potential as well.

\textcolor{black}{Another possible reason for the decrease of the membrane potential is due to incomplete mixing in the reservoir and non-ideality of the electrolyte solution at high concentration, which are out of the scope of this paper.}


\subsection{Combination of electronic and chemical charge} \label{sec:combine}
Let us now consider the case when the total surface charge results from the presence of both electronic and chemical charge. Although they are not directly coupled, we have to determine them consistently since they are correlated by the the potential distribution in the pore.

The factor $\Ndl/\cs$ is used to characterize the ratio between the electronic and the chemical charge, where $\Ndl$ is the dimensionless number of surface sites for chemical charge, and $\cs$ is the dimensionless Stern layer capacitance. In fact, the amount of the chemical charge can be characterized by the maximum possible charge density, i.e., $N \, e$, while the amount of the electronic charge can be characterized by the product of the capacitance $\Cs$ and the characteristic potential $ R_g T/F$. The ratio between the two gives $\Ndl/\cs$.

\begin{figure*}[!htb]
    \begin{minipage}[t]{0.5\textwidth}
    \includegraphics[scale=0.7]{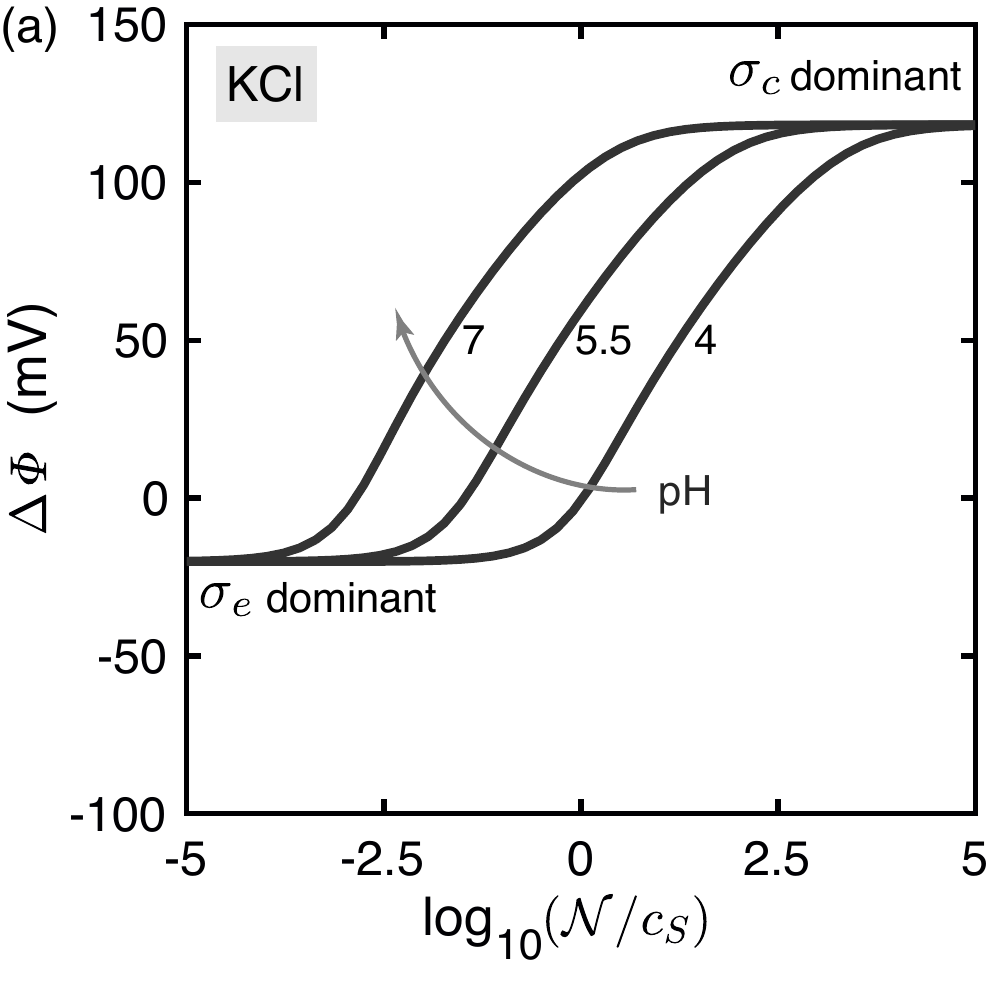}
    \end{minipage}%
    \begin{minipage}[t]{0.5\textwidth}
    \includegraphics[scale=0.7]{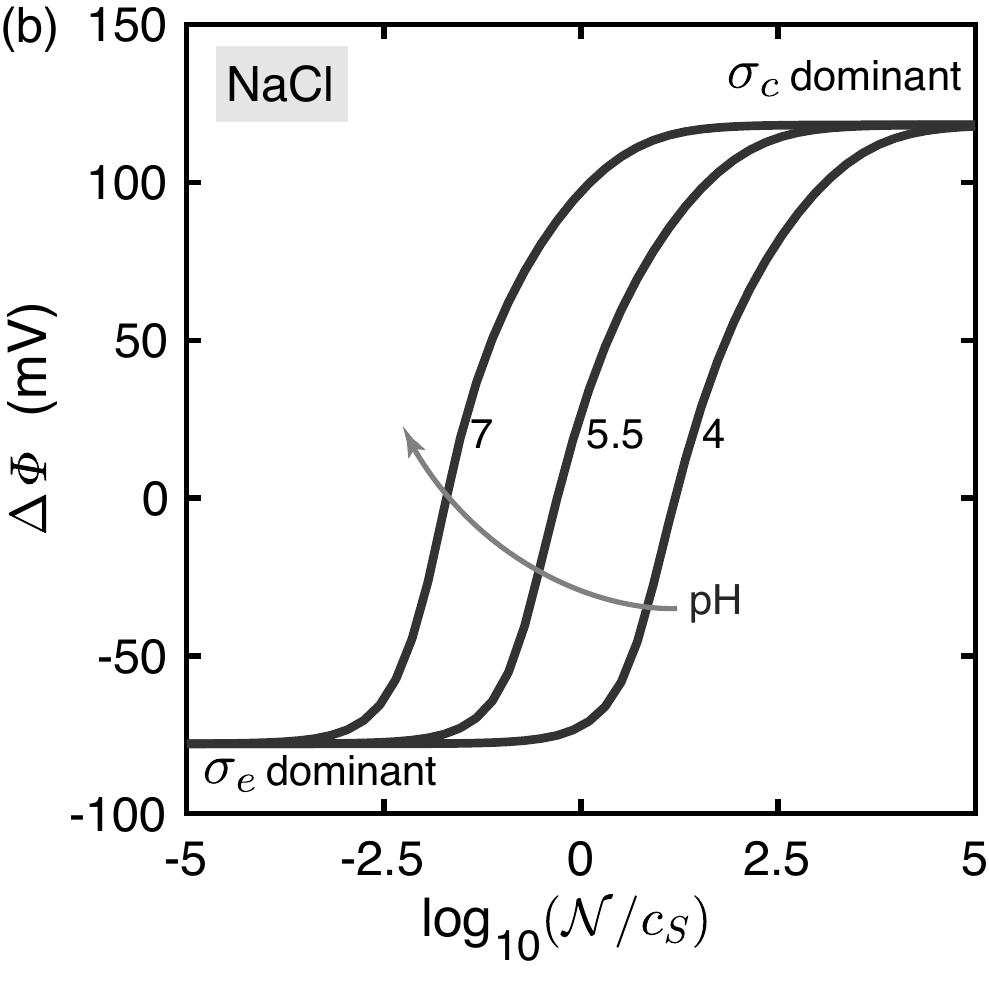}
    \end{minipage}%
    \caption{Membrane potential with both electronic and chemical charge for different $\Ndl/\cs$ ratios at zero current with $\Cs$ fixed at $\SI{1}{F/m^2}$. $\Ndl$ is the dimensionless number of surface sites for chemical charge, and $\cs$ is the dimensionless Stern layer capacitance ($C_h = \SI{10}{mM}$, $ C_l = \SI{0.1}{mM}$, $\mathrm{pK} = 7.5$ and $\overline{X}_e = 0$).}
    \label{fig:Dphi_both}
\end{figure*}

Figure \ref{fig:Dphi_both} shows the change of the membrane potential with $\Ndl/\cs$ for different pH. In general, if $\Ndl \gg \cs$, the chemical charge will dominate and become the main mechanism for membrane potential generation (Fig.~\ref{fig:sketch}c), while if $\Ndl \ll \cs$, the electronic charge dominates and the charging mechanism follows that shown in Fig.~\ref{fig:sketch}b. In the chemical-charge-dominated regime, the difference in ion mobilities giving rise to the diffusion potential plays minor role, so that the membrane potential is the same for both \ce{KCl} and \ce{NaCl}. However, the difference in mobility, determining the enhancement effect, strongly influences the membrane potential in the electronic-charge-dominated regime. In addition, due to the pH-dependence of chemical charge, the transition will shift towards the chemical-charge-dominated regime if the reservoir pH increases.

\subsection{Energy generation from concentration difference} \label{sec:energy_generation}
\begin{figure}
    \centering
    \begin{minipage}[t]{\textwidth}
    \includegraphics[scale=0.7]{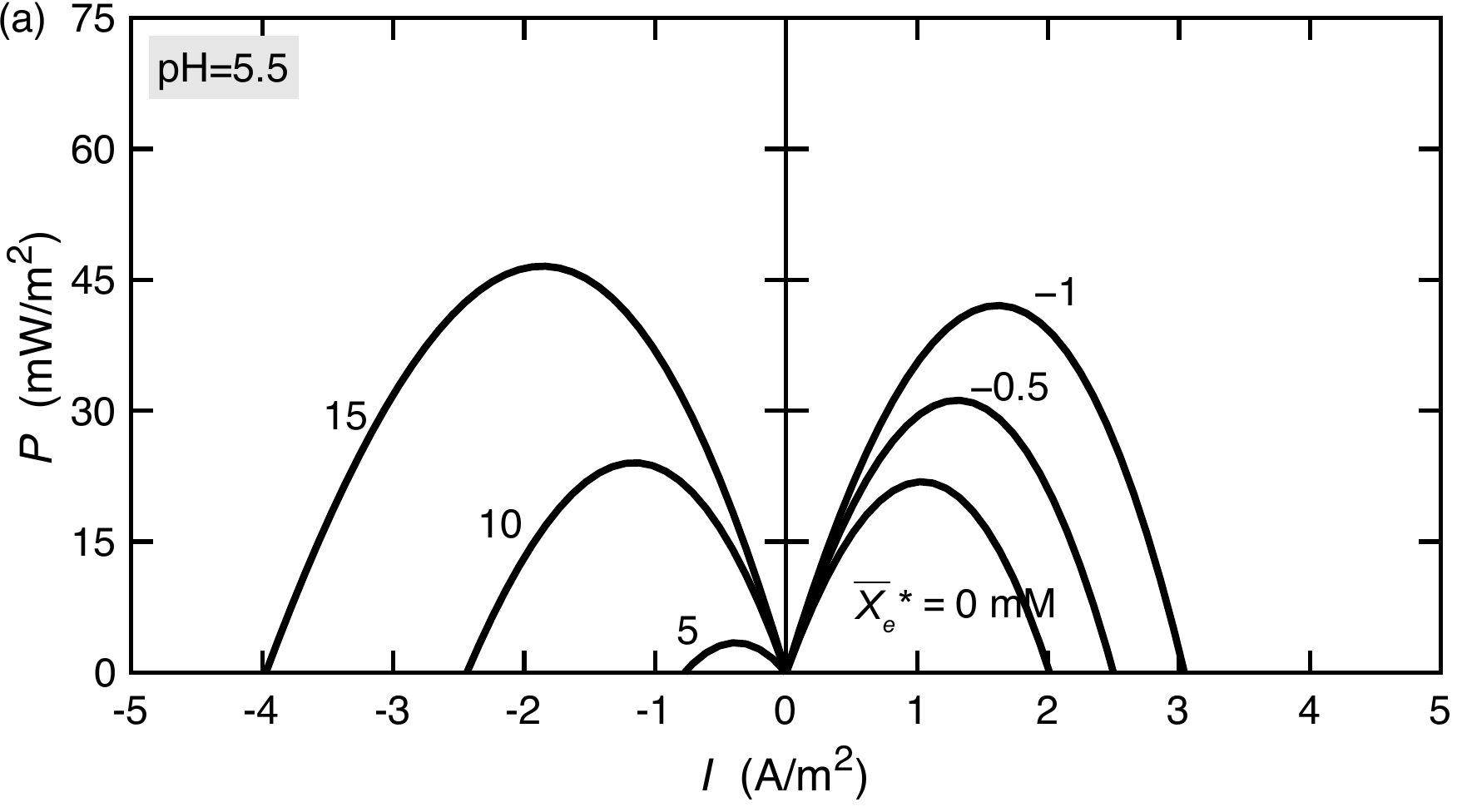}
    \end{minipage}
    \begin{minipage}[t]{\textwidth}
    \includegraphics[scale=0.7]{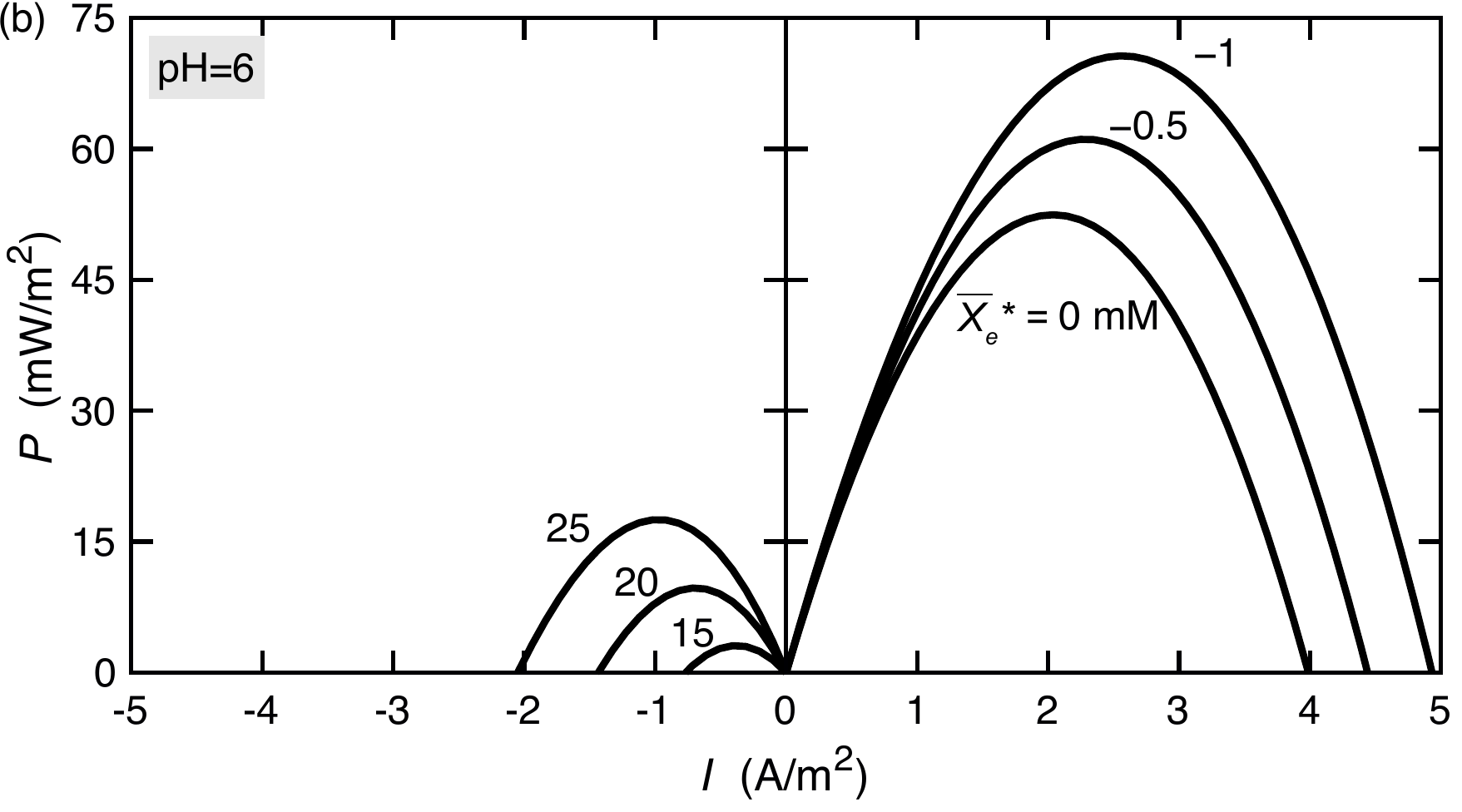}
    \end{minipage}
    \caption{Osmotic power density generated from a concentration difference ($C_h = \SI{1}{mM}$, $C_l = \SI{0.1}{mM}$) by an $\SI{8}{nm}$-radius pore filled of \ce{NaCl} at (a) $\mathrm{pH} = 5.5$ and (b) $\mathrm{pH} = 6$. The pore bears both electronic charge ($\Cs = \SI{1}{F/m^2}$) and pH-dependent chemical charge ($N = \SI{1}{nm^{-2}}$, $\mathrm{pK} = 7.5$). A different electronic charge is supplied or withdrawn to reach a certain total volumetric electronic charge density $\overline{X}^*_e$.}
    \label{fig:P_Xe}
\end{figure}

If connected to an external load, a permselective membrane can convert the osmotic energy from a concentration difference to electrical energy. If the membranes are stacked in a way of alternating permselectivity, it forms the process of reverse electrodialysis~\cite{Fair1971, Post2007, Tedesco2016, Siria2017}. The power density of this conversion is the product of the electric current density and the potential difference across the membrane, which is the membrane potential at open-circuit condition, i.e., zero electric current. The power density reaches zero at both open-circuit condition and short-circuit condition, where the maximum electric current is reached. It typically follows a parabola profile with the electric current, and the maximum power density is achieved at around the half of the maximum current density (Fig.~\ref{fig:P_Xe}). \textcolor{black}{For a negatively charged membrane (i.e., cation-selective), the cation is the main charge carrier and the energy conversion process is operated at a positive current (right branch of Fig.~\ref{fig:P_Xe}). Extra negative electronic charge can be supplied to the pore wall to improve the performance of the membrane. As shown in Fig.~\ref{fig:P_Xe}a, the maximum power density almost doubles from $\SI{22}{mW/m^2}$ to $\SI{42}{mW/m^2}$ when the extra volumetric electronic charge density $\overline{X}_e^* = \overline{X}_e \, C_0$ reaches $\SI{-1}{mM}$ in the pore volume, equivalently, about $\SI{-0.38}{mC/m^2}$. At the same time, the corresponding current density for the maximum power density shifts from $\SI{1.06}{A/m^2}$ to $\SI{1.63}{A/m^2}$. In contrast, if electrons are withdrawn from the pore, the membrane becomes positively charged (i.e., anion-selective) and the direction of the electric current is reversed to generate power (left branch of Fig.~\ref{fig:P_Xe}). Because of the negative chemical charge, it requires more electronic charge to reach same power density for this case.}

The $\Ndl/\cs$ ratio in Fig~\ref{fig:P_Xe} is about 6.2. In this case, according to Fig.~\ref{fig:Dphi_both}b, the membrane potential is very sensitive to the change of pH. A slight increase of pH to 6 gives rise to more negative chemical charge and makes the power density higher when $I > 0$. \textcolor{black}{At the same time, it requires further withdrawal of electrons in comparison to the case of pH=5.5 to overcome the chemical charge and change the polarity of the membrane.} Note that the concentrations used in Fig.~\ref{fig:P_Xe} ($C_h = \SI{1}{mM}$, $C_l = \SI{0.1}{mM}$) are relatively low, so only a small amount of extra electronic charge supplied makes large impact on the power density. At a higher salt concentration, however, more electronic charge is required to increase the power density.

\begin{figure}
    \centering
    \begin{minipage}[t]{\textwidth}
    \includegraphics[scale=0.7]{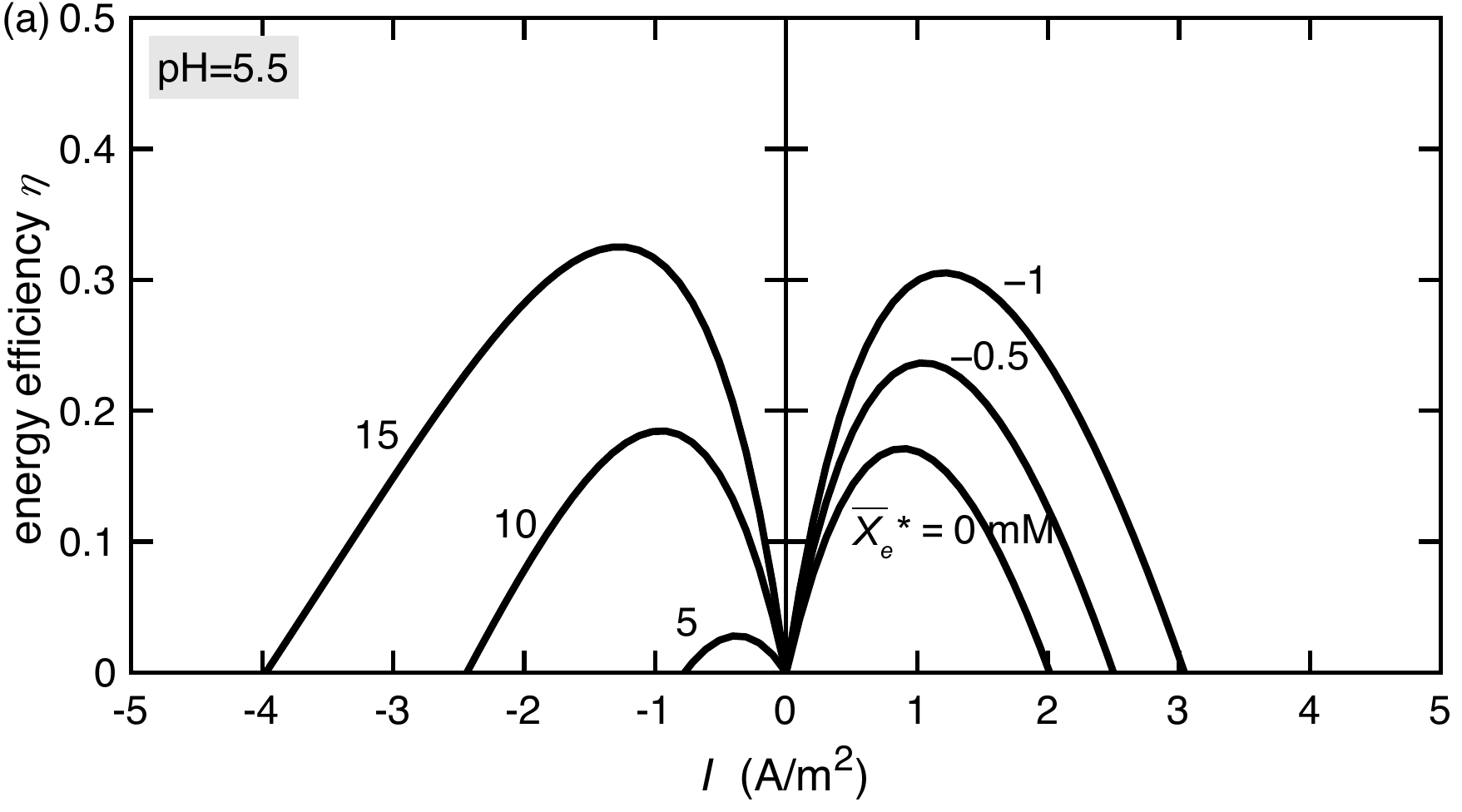}
    \end{minipage}
    \begin{minipage}[t]{\textwidth}
    \includegraphics[scale=0.7]{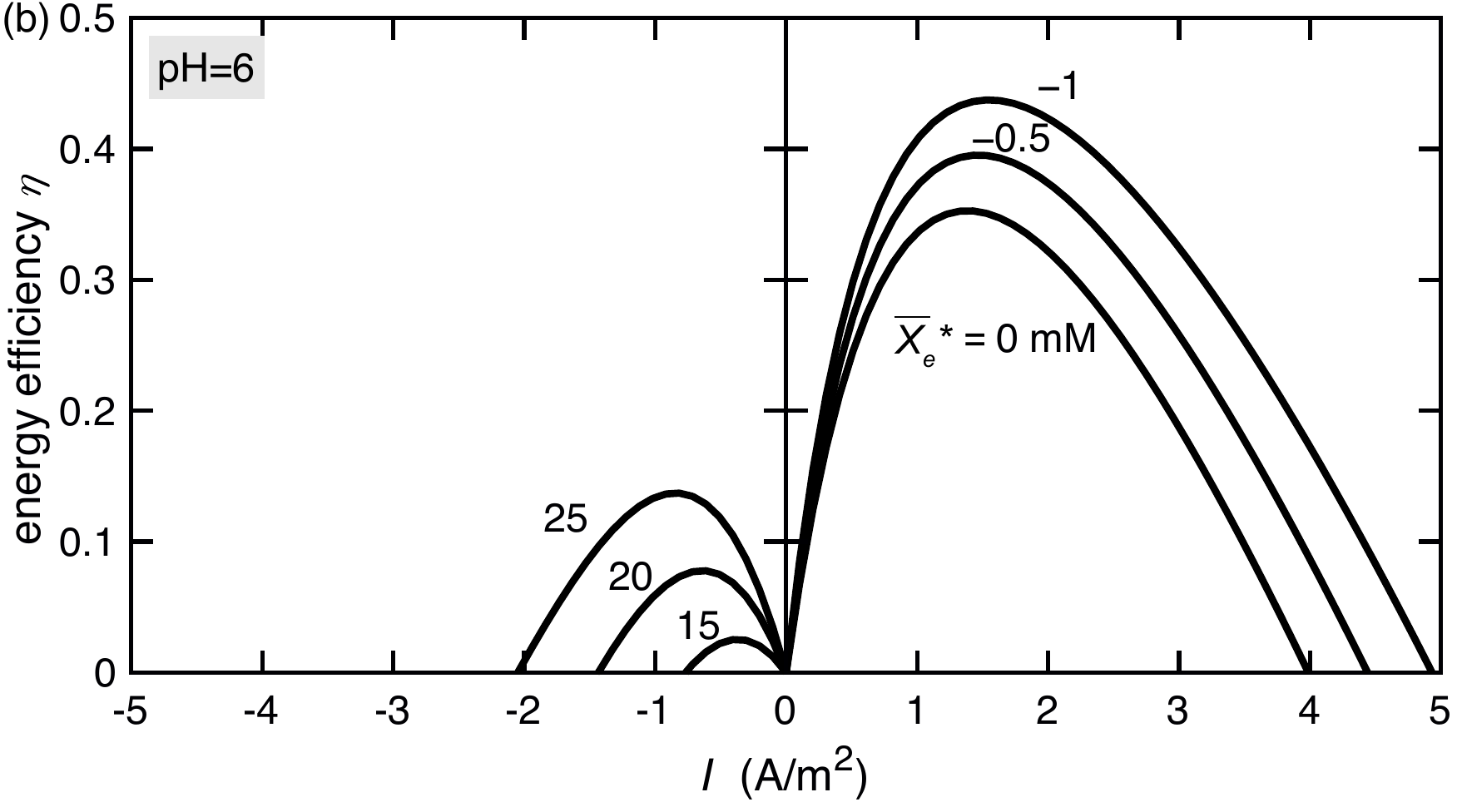}
    \end{minipage}    
    \caption{Energy efficiency of osmotic power generated from a concentration difference ($C_h = \SI{1}{mM}$, $C_l = \SI{0.1}{mM}$) by an $\SI{8}{nm}$-radius pore filled of \ce{NaCl} at (a) $\mathrm{pH} = 5.5$ and (b) $\mathrm{pH} = 6$. The pore bears both electronic charge ($\Cs = \SI{1}{F/m^2}$) and pH-dependent chemical charge ($N = \SI{1}{nm^{-2}}$, $\mathrm{pK} = 7.5$). A different electronic charge is supplied or withdrawn to reach a certain total volumetric electronic charge density $\overline{X}^*_e$.}
    \label{fig:eta_Xe}
\end{figure}

The overall energy efficiency of this conversion process is defined as~\cite{Gross1968, Fair1971, Peters2016}
\begin{equation}
    \eta = \frac{j_\mathrm{ch} \, \Delta \phi}{j_\mathrm{ions} \ln(c_h/c_l) - 2 u (c_h - c_l)},
\end{equation}
which is the ratio of the generated electrical power to the Gibbs free energy of mixing taking into account of the adverse effect of advection. It measures the effectiveness of the membrane in overcoming the dissipation effect by entropy generation. Note that the energy consumption by supplying the extra electronic charge is not considered, since the charge density is fixed in the energy generation process if there is no electron leakage by Faradaic reactions. Like the power density, the energy efficiency increases with the addition of negative electronic charge. \textcolor{black}{Energy efficiency in Fig.~\ref{fig:eta_Xe} follows a similar trend as the power density in Fig.~\ref{fig:P_Xe}. When operated at positive current, the energy efficiency reaches 30\% for a current density of $\SI{1.2}{A/m^2}$ at pH = 5.5, and 44\% for a current density of $\SI{1.5}{A/m^2}$ at pH = 6 with an extra total volumetric electronic charge density of $\SI{-1}{mM}$. In contrast, when charged positively and operated at negative current, the energy efficiency declines to less than 5\% with $\overline{X}^*_e=\SI{5}{mM}$ at pH=5.5 and $\overline{X}^*_e=\SI{15}{mM}$ at pH=6.}

\begin{figure}
    \centering
    \begin{minipage}[t]{0.5\textwidth}
    \includegraphics[scale=0.7]{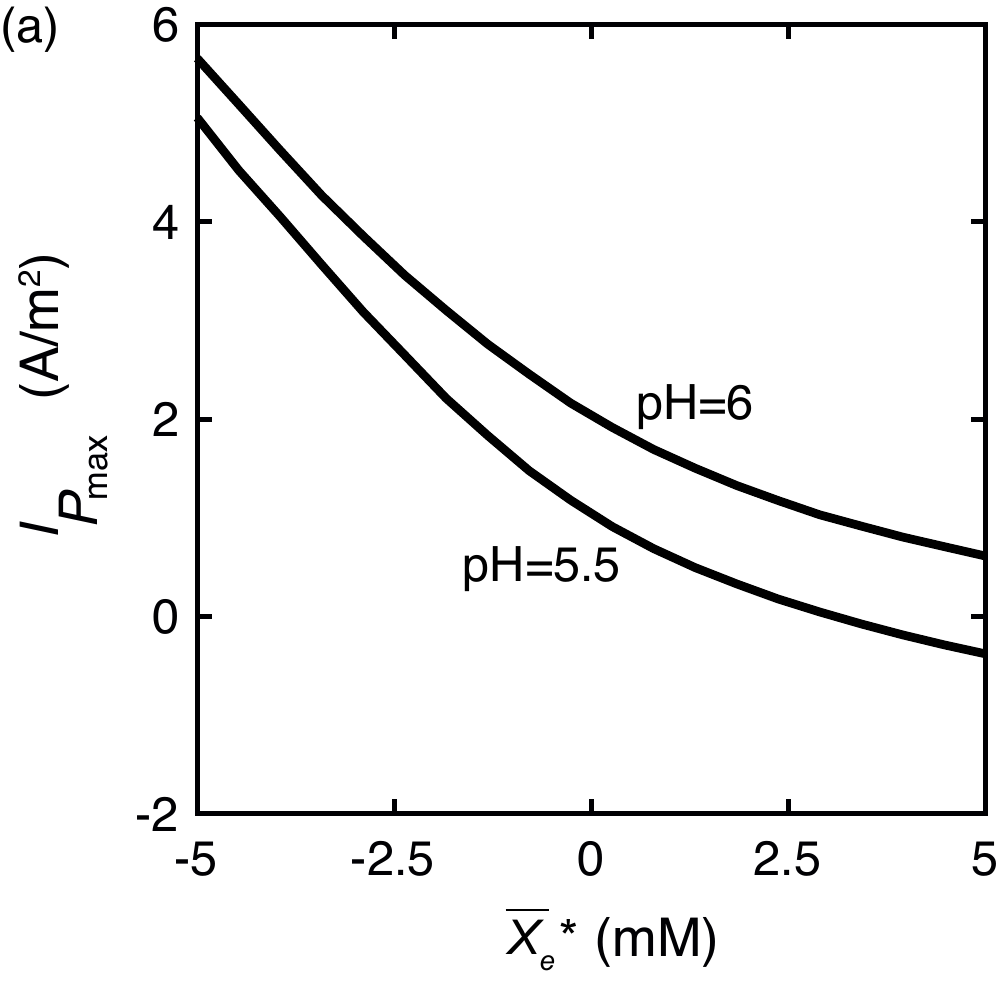}
    \end{minipage}%
    \begin{minipage}[t]{0.5\textwidth}
    \includegraphics[scale=0.7]{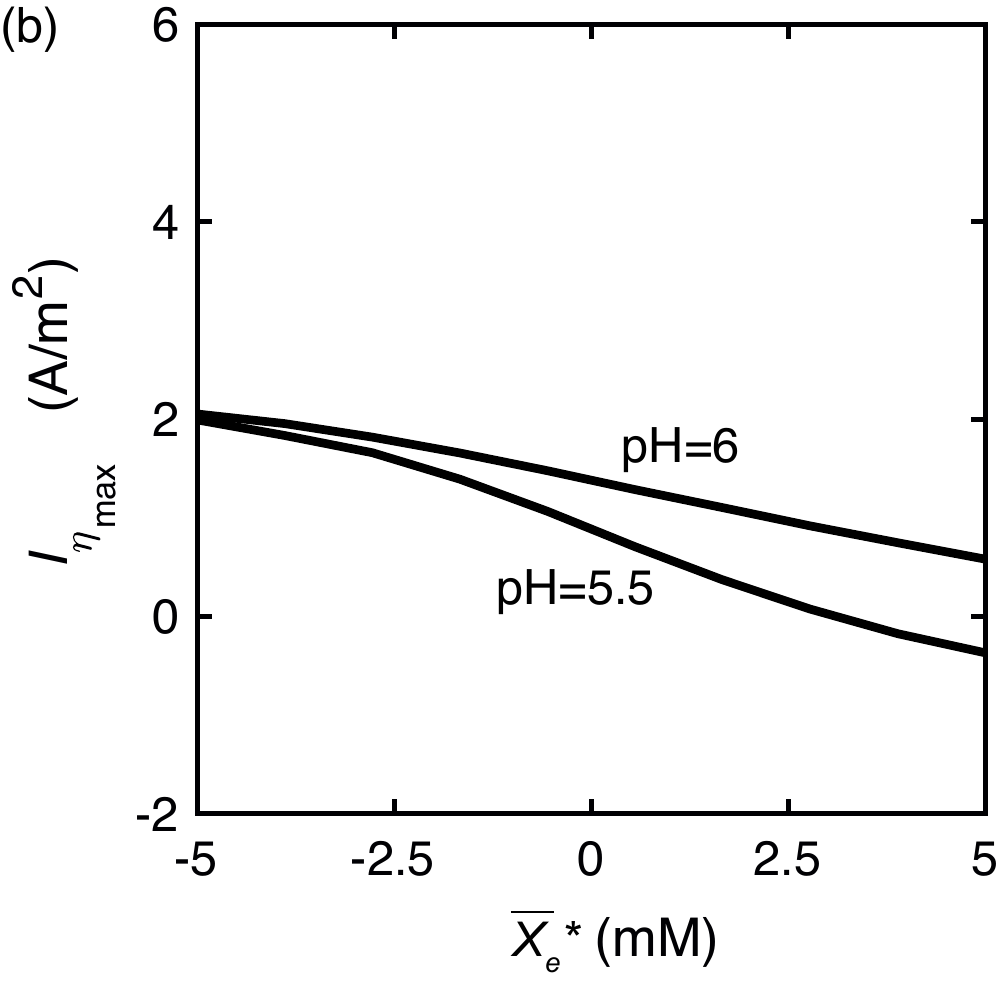}
    \end{minipage}    
    \caption{Optimum current density at (a) maximum power density and (b) maximum energy efficiency with different total electronic charge density $\overline{X}^*_e$. $H_p = \SI{8}{nm}$, $C_h = \SI{1}{mM}$, $C_l = \SI{0.1}{mM}$, $\Cs = \SI{1}{F/m^2}$, $N = \SI{1}{nm^{-2}}$, $\mathrm{pK} = 7.5$.}
    \label{fig:Imax_Xe}
\end{figure}

\textcolor{black}{Figure \ref{fig:Imax_Xe} shows the optimum current density to reach the maximum power density and energy efficiency when the pore is charged with different total electronic charge density $\overline{X}^*_e$. While the optimum current density for power density $I_{P_\mathrm{max}}$ is sensitive to a change in $\overline{X}^*_e$, the optimum current density for energy efficiency $I_{\eta_\mathrm{max}}$ is much less dependent on $\overline{X}^*_e$. A good operation condition for the current density may be in between of these two values $I_{P_\mathrm{max}}$ and $I_{\eta_\mathrm{max}}$.}

\section{Conclusions}
In this work, we have developed a theoretical description of ion transport in nanoporous membranes in the presence of both electronic and chemical charge on the pore surface. The former is induced by the electrons in the conductive pore surface, while the latter originates from ionization of surface chemical groups. Even if no external charge is injected, the electronic charge can redistribute along the pore, leading to intriguing profiles in ion concentration and potential (see Fig.~\ref{fig:profile_Xe}). The pH-dependent chemical charge is regulated by proton transport, giving rise to a potential difference within the membrane. When both are present, the two types of surface charge are correlated by the potential distribution in the pore.

The electrical potential across the membrane is investigated at a zero electric current condition. The electronic charge is found to strongly enhance the diffusion potential through redistribution of electrons, even for \ce{KCl} with a minor difference in diffusion coefficients, while the pH-dependent chemical charge leads to an increase of the electrical potential within the membrane if the pH in both reservoirs is kept the same. Our one-dimensional model shows good agreement with both experimental data and results of two-dimensional space charge model as long as the pore size is relatively small compared with the Debye length. In addition, the performance of the membrane used for energy conversion from a concentration difference is also investigated for non-zero electric currents. By tuning electronic charge of the membrane, the selectivity of the membrane can be controlled. For example, if extra negative(positive) charge is supplied to the cation-selective(anion-selective) membrane, the power density and energy efficiency for RED can be improved. This flexible control of the membrane selectivity may open opportunities for new designs in RED and other relevant applications.

\begin{acknowledgments}
This work was performed in the cooperation framework of Wetsus, European Centre of Excellence for Sustainable Water Technology (www.wetsus.eu). Wetsus is co-funded by the Dutch Ministry of Economic Affairs and Ministry of Infrastructure and Environment, the Province of Frysl{\^a}n, and the Northern Netherlands Provinces. LZ and PMB thank the participants of the research theme Advanced Water Treatment for fruitful discussions and financial support. IR acknowledges the support of Russian Foundation for Basic Research, Government of Krasnoyarsk Territory, Krasnoyarsk Regional Fund of Science in the frame of research project 18--48--242011 "Mathematical modelling of synthesis and ionic transport properties of conductive nanoporous membranes". IR also acknowledges the Dutch Science Foundation NWO for a Visitors travel grant 040.11.694.
\end{acknowledgments}

\bibliography{ref}

\end{document}